\documentclass[12pt]{article}

\usepackage{epsfig}
\usepackage{amsmath}
\usepackage{amssymb}
\def\permil{\%\raise.10ex\hbox{$_{\scriptstyle 0}$}}
\def\be{\begin{equation}}
\def\ee{\end{equation}}
\def\bea{\begin{eqnarray}}
\def\eea{\end{eqnarray}}
\def\bsp{\begin{split}}
\def\esp{\end{split}}

\def\cdott{\!\cdot\!}

\def\cVvv{{\cal V}_{LON_c}^{vl\{a'\},\{b\}}}

\def\B{{\cal C}}
\def\cVr{{\cal V}_{LON_c}^{r\{a'\}\{b\}}}
\def\cVrrrr{{\cal V}_{LON_c}^{r\{a'\},\{b\},\tau=-2}}
\def\cVrr{{\cal V}_{LON_c}^{r\{a'\}}}
\def\cVrrr{{\cal V}_{subN_c}^{r\{a'\},\{b\}}}

\def\csVrrr{{\cal V}_{subN_c}^{r\{a'\},\{b\},\tau=4}}

\def\imp{\phi}

\def\impfa{\imp_{\{a'\}}\otimes}
\def\verta{{\cal V}^{a'_1a'_2;a_1a_2a_3a_4}}

\def\ak{|k|}
\def\ak_1{|k_1|}
\def\ak_2{|k_2|}
\def\ak_3{|k_3|}
\def\ak_4{|k_4|}
\def\aw{|w|}
\def\aw_1{|w_1|}
\def\aw_2{|w_2|}
\def\aw_3{|w_3|}
\def\aw_4{|w_4|}

\def\k{\textbf{k}}
\def\l{\textbf{l}}
\def\q{\textbf{q}}
\def\m{\textbf{m}}
\def\r{\textbf{r}}
\def\w{\textbf{w}}
\def\p{\textbf{p}}
\def\x{\textbf{x}}
\def\b{\textbf{b}}
\def\kpp{\mbox{\boldmath{$\kappa$}}}

\newcommand\epjc[3]{{\it Eur. Phys. J.}{\bf C #1} (#2) #3}  
\newcommand\plb[3]{{\it Phys. Lett. }{\bf B #1} (#2) #3}     
\begin{document}
\titlepage
\begin{flushright}
DESY-07-185 \\
October 2007
\end{flushright}
\vspace*{1in}
\begin{center}
{\Large \bf A Momentum Space Analysis\\[0.5cm]
 of the Triple Pomeron Vertex in pQCD}\\
\vspace*{0.5cm}
J. \ Bartels$^{(a)}$
and K.\ Kutak$^{(b)}$ \\
\vspace*{0.5cm}
 $^{(a)}${\it II.\ Institut f\"ur Theoretische Physik,
    Universit\"at Hamburg \\ Luruper Chaussee 149, Germany}\\
$^{(b)}$ {\it DESY Notkestrasse 85, 22607 Hamburg, Germany}\\

\end{center}
\vspace*{1cm}
\centerline{(\today)}

\vskip1cm
\begin{abstract}
\noindent
We study properties of the momentum space Triple Pomeron Vertex in 
perturbative QCD. Particular attention is given to the collinear limit where 
transverse momenta on one side of the vertex are much larger than on 
the other side. We also comment on the kernels in nonlinear evolution equations.
\end{abstract}

\section{Introduction}

The Triple Pomeron Vertex (TPV) in perturbative QCD \cite{BB,BW,Balvertex} 
has attracted significant 
attention in recent years. It is derived from the $2 \to 4$ transition vertex 
in QCD reggeon field theory which represents the high energy description 
of QCD. In recent years particular interest has come from studies of 
nonlinear evolution 
equations, e.g. the Balitsky-Kovchegov equation \cite{Balvertex,Bal,Kov},   
where the nonlinearity is given by the TPV. More recently, 
also generalizations of the nonlinear evolution have been considered 
\cite{Loop} which contain pomeron loops \cite{LoopBart}. 
Again, the TPV plays a central role in these investigations. 
Whereas in many studies and applications it is convenient to use the 
coordinate representation, it is important to understand the structure 
also in momentum space.        
   
In this paper we will investigate some aspects of the TPV, 
starting from the momentum space representation of the $2 \to 4$ gluon 
transition vertex, from which the TPV vertex has originally been derived 
\cite{BB, BW}.
Many of the studies of the nonlinear evolution equations have been 
done in the context of deep inelastic scattering where a virtual photon 
scatters off a single nucleon or off a nucleus. In both cases the momentum 
scale of the photon is much larger that the typical scale of the hadron or 
nucleus, i.e. one is dealing with asymmetric momentum configurations.        
As a first step of investigating the TPV, therefore, we will focus on 
investigating the limit where the transverse momenta are strongly ordered. 
We also review and discuss the nonlinear evolution equation that have been proposed in the literature 
\cite{KKM01,KK03}, and we comment on the use of a twist-expansion in the 
low-$x$ limit. 

The paper is organized as follows. In the following section 2 we define the setup of
our calculation, and we construct the elastic amplitude for photon-photon 
scattering 
with exchange of a four gluon BKP state. In the large-$N_c$ limit, this 
reduces to photon-photon scattering with the exchange of a pomeron loop. 
In the remaining part of
section 2 we define the Mellin transform of a pomeron loop and specify our 
use of 'collinear' and 'anticollinear limits'. In section 3 we study
the collinear  limit of the Triple Pomeron Vertex in the large $N_c$ limit. 
Section 4 contains results of the 
analysis of the anticollinear limit in the large $N_c$ limit. 
In section 5 we extend the analysis to finite $N_c$. In section 6 we derive 
a hierarchy of nonlinear evolution equations which describe the 
interaction of a photon with a hadronic target. We also show that, in the mean
field approximation, we obtain a nonlinear evolution equation for the 
unintegrated gluon density. Section 7 contains a few comments on the relation 
of this equation with other nonlinear evolution equations described in the literature. 
We end the paper with a few conclusions.

\section{The $2 \to 4$ gluon transition vertex}
The LO momentum space expression for the $2 \to 4$ gluon transition vertex 
has been derived in connection with the diffractive dissociation of the 
virtual photon in deep inelastic electron proton scattering \cite{BW}.
\begin{figure}[h!]
\centerline{\epsfig{file=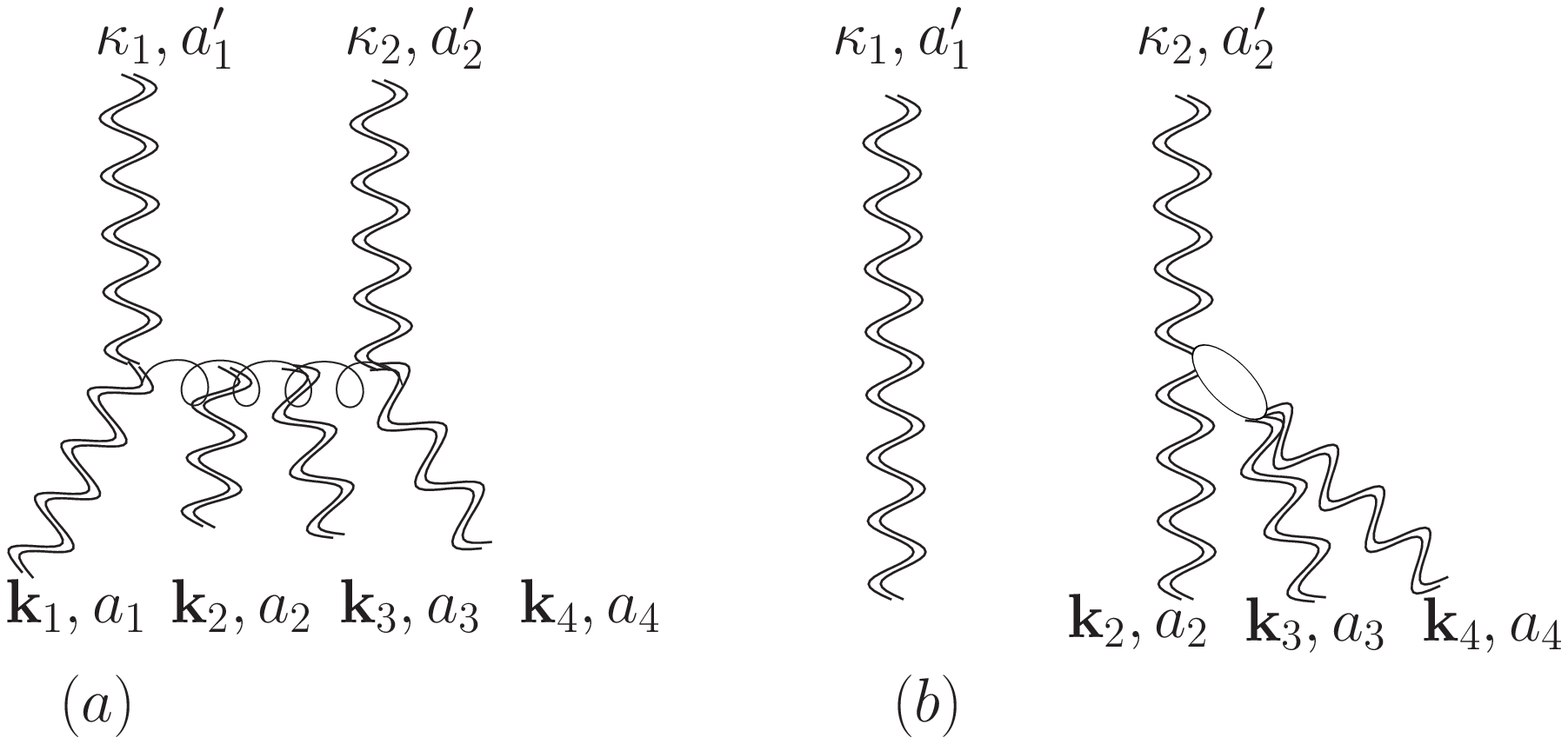,height=4cm}}
\caption{\em Examples of diagrams that contribute to the $2 \to 4$ gluon 
transition vertex (wavy vertical lines represent reggeized gluons): 
(a) real emission, (b) a disconnected contribution.}
\label{fig:diagramy2}
\end{figure}
More precisely, the process  $\gamma^* +q \to (q\bar{q} + n \;\;gluons) +q$ 
has been investigated in the triple Regge limit. The resulting vertex  
consists of three pieces (we follow the notation of \cite{BLV}):
\be
\begin{split}
\verta(\kpp_1,\kpp_2;\k_1,\k_2,\k_3,\k_4)=\frac{\sqrt{2}\pi\delta^{a'_1a'_2}}{N_c^2-1}
\Bigg[\delta^{a_1a_2}\delta^{a_3a_4}V(\kpp_1,\kpp_2,\k_1,\k_2,\k_3,\k_4)\\+
\delta^{a_1a_3}\delta^{a_2a_4}V(\kpp_1,\kpp_2,\k_1,\k_3,\k_2,\k_4)
+\delta^{a_1a_4}\delta^{a_2a_3}V(\kpp_1,\kpp_2,\k_1,\k_4,\k_2,\k_3)\Bigg],
\label{eq:fvertex}
\end{split}
\ee
where $\kpp_1+\kpp_2 = \k_1+\k_2+\k_3+\k_4=\q$,
and the subscripts $a_i'$, $a_i$ refer to the color degrees of freedom of the
reggeized gluons. It is convenient to express the 'basic vertex function'
$V(\kpp_1,\kpp_2,\k_1,\k_2,\k_3,\k_4)$ in terms of another function
$G(\kpp_1,\kpp_2,\k_1,\k_2,\k_3)$: 
\be
\begin{split}
V(\kpp_1,\kpp_2,\k_1,\k_2,\k_3,\k_4)=
\frac{1}{2}g^4\bigg[
G(\kpp_1,\kpp_2,\k_1,\k_2+\k_3,\k_4)+G(\kpp_1,\kpp_2,\k_2,\k_1+\k_3,\k_4)\\
+G(\kpp_1,\kpp_2,\k_1,\k_2+\k_4,\k_3)
+G(\kpp_1,\kpp_2,\k_2,\k_1+\k_4,\k_3)-G(\kpp_1,\kpp_2,\k_1+\k_2,\k_3,\k_4)\\
-G(\kpp_1,\kpp_2,\k_1+\k_2,\k_4,\k_3)
-G(\kpp_1,\kpp_2,\k_1,\k_2,\k_3+\k_4)-G(\kpp_1,\kpp_2,\k_2,\k_1,\k_3+\k_4)\\
+G(\kpp_1,\kpp_2,\k_1+\k_2,-,\k_3+\k_4)\bigg].
\label{eq:vertex}
\end{split}
\ee
This function $G(\kpp_1,\kpp_2,\k_1,\k_2,\k_3)$ \cite{BV,JPV} generalizes
the $G$ function introduced  in \cite{BW} to the non-forward direction. This function can again
be split into two pieces:
\be
G(\kpp_1,\kpp_2,\k_1,\k_2,\k_3)=G_1(\kpp_1,\kpp_2,\k_1,\k_2,\k_3)
+G_2(\kpp_1,\kpp_2,\k_1,\k_2,\k_3),
\label{eq:funkg1}
\ee
where the first part contains the 'connected contributions' (also: 
'real contributions'):
\bea
G_1(\kpp_1,\kpp_2,\k_1,\k_2,\k_3)=
\frac{(\k_2+\k_3)^2\kpp_1^2}{(\kpp_1-\k_1)^2}+\frac{(\k_1+\k_2)^2\kpp_2^2}{(\kpp_2-\k_3)^2}-
\frac{\k_2^2\kpp_1^2\kpp_2^2}{(\kpp_1-\k_1)^2(\kpp_2-\k_3)^2}\nonumber\\
-(\k_1+\k_2+\k_3)^2,
\label{eq:funkg2}
\eea
and the second one takes care of the disconnected ('virtual') pieces:
\bea
g^2 G_2(\kpp_1,\kpp_2,\k_1,\k_2,\k_3)&=&-\frac{\kpp_1^2\kpp_2^2}{N_c}\Big([\omega(\k_2)-
\omega(\k_2+\k_3)]\delta^{(2)}(\kpp_1-\k_1)
\nonumber\\&&+[\omega(\k_2)-\omega(\k_1+\k_2)]\delta^{(2)}
(\kpp_1-\k_1-\k_2)\Big).
\label{eq:funkg3}
\eea
Here $\omega(\k)$ denotes the trajectory function.:
\be
\omega(\k)=-N_c g^2\int\frac{d^2\l}{(2\pi)^3}\frac{\k^2}{\l^2+(\k-\l)^2}\frac{1}{(\k-\l)^2}.
\ee
The vertex 
(\ref{eq:fvertex}) is completely symmetric under the permutation of the four
gluons. It is infrared finite, it has been shown to be invariant 
under M\"obius transformations ~\cite{BLW}, and it vanishes when $\kpp_i$ or $\k_i$
goes to zero.

\begin{figure}[t!]
\centerline{\epsfig{file=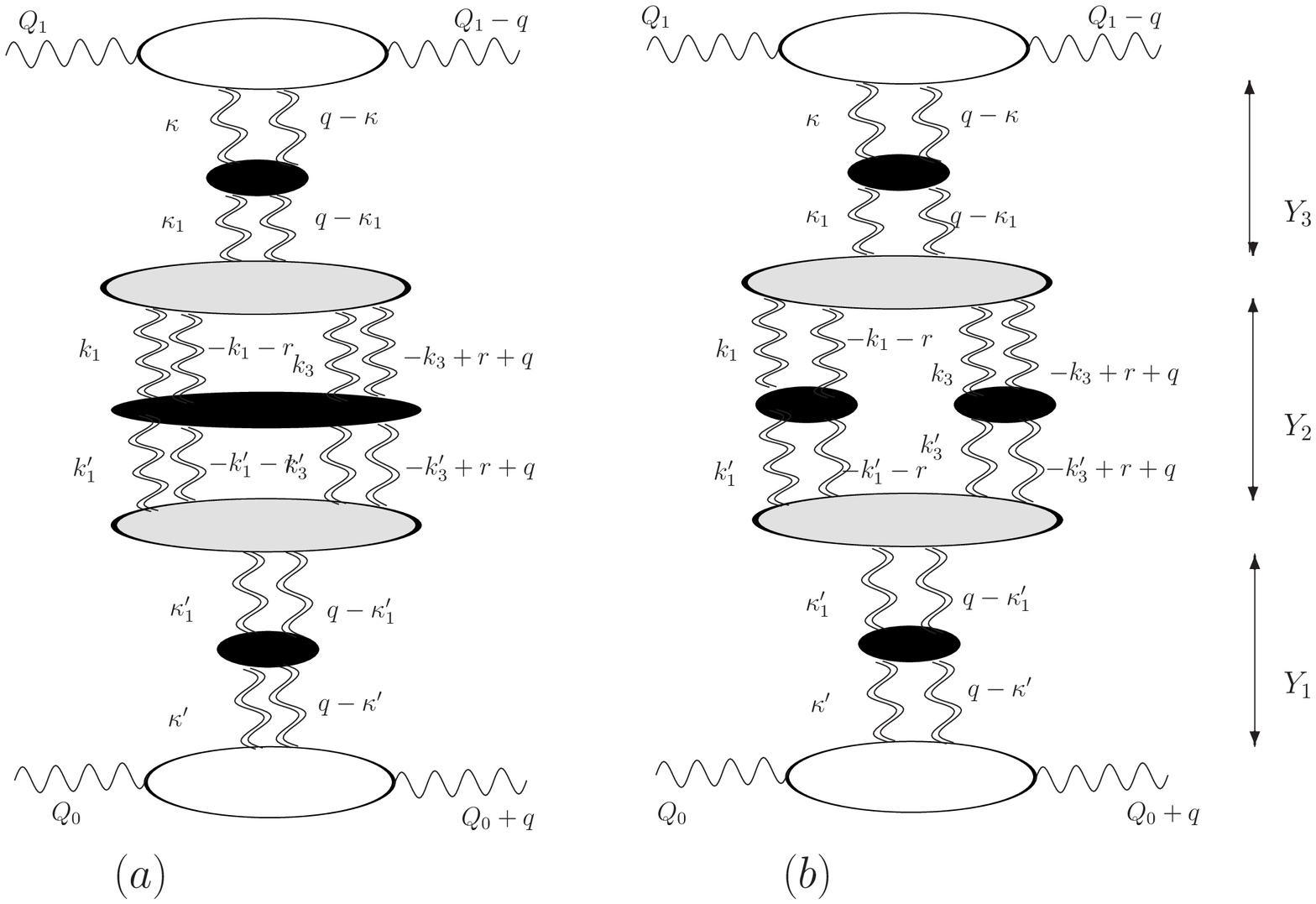,height=10cm}}
\caption{\em Contributions to the elastic scattering of two virtual photons
which contain the $2 \to 4$ gluon vertex (dark blobs represent Green's 
functions of reggeized gluons): 
(a) the four gluon BKP state; (b) a Pomeron loop.}
\label{fig:pomloopBKP3}
\end{figure}
This vertex can be used to construct, in reggeon field theory, 
the selfenergy $\Sigma$, of the BFKL Green's function 
(Fig. \ref{fig:pomloopBKP3}). In the lowest order contribution to $\Sigma$, 
we have a BKP state between two $2 \to 4$ vertices, which contains    
all pairwise interactions of four reggeized $t$-channel
gluons. Its Green's function satisfies the following evolution equation:
\bea
(\omega -\omega(\k_1)-\omega(\k_2)-\omega(\k_3)-\omega(\k_4))
{\cal G}^{(4)\;\{a_i\},\{a'_i\}}_{\omega} (\{\k_i\},\{\k'_i\})=&\nonumber\\
{\cal G}^{(4)0\;\{a_i\},\{a'_i\}}(\{\k_i\},\{\k'_i\})\;
+ \; \sum_{(ij)} \frac{1}{\k_i^2 \k_j^2}
K_{2\rightarrow 2}^{\{a\}\rightarrow\{b\}}\,\otimes\,
{\cal G}^{(4)\;\{b_i\}\{a'_i\}}_{\omega} (\{\k_i\}\{\k'_i\}),
&
\eea
where we have used the shorthand notation $\{\k_i\}=(\k_1,\k_2,\k_3,\k_4)$ etc.
The sum extends over all pairs $(ij)$ of gluons, the kernel
$ K_{2\rightarrow 2}^{\{a\}\rightarrow\{b\}}$ includes the color tensor
$f_{a_i b_i c} f_{a_j b_j c}$
\be
K_{2\rightarrow 2}^{\{a_i\}\rightarrow\{b_i\}}=
g^2\; f_{b_1a_1c}f_{ca_2b_2}\left[\r^2-\frac{\k_1^2(\k-\r)^2}{(\k_1-\k)^2}-\frac{\k^2(\k_1-\r)^2}{(\k_1-\k)^2}\right],\\
\label{eq:2to2kernel}
\ee
and the convolution symbol $\otimes$ stands for
$\int\!\!\frac{d\k^2}{(2\pi)^3}$.
The inhomogeneous term has the form:
\bea
\delta^{(2)}(\sum \k_i-\sum \k_i') {\cal G}^{(4)0\;\{a_i\}\{a'_i\}}
(\{\k_i\},\{\k'_i\})= (2 \pi)^9 \prod_1^4
\frac{\delta_{a_i a'_i} \delta^{(2)}(\k_i-\k'_i)}{\k_i^2} .
\label{G40}
\eea
\begin{figure}[t!]
\centerline{\epsfig{file=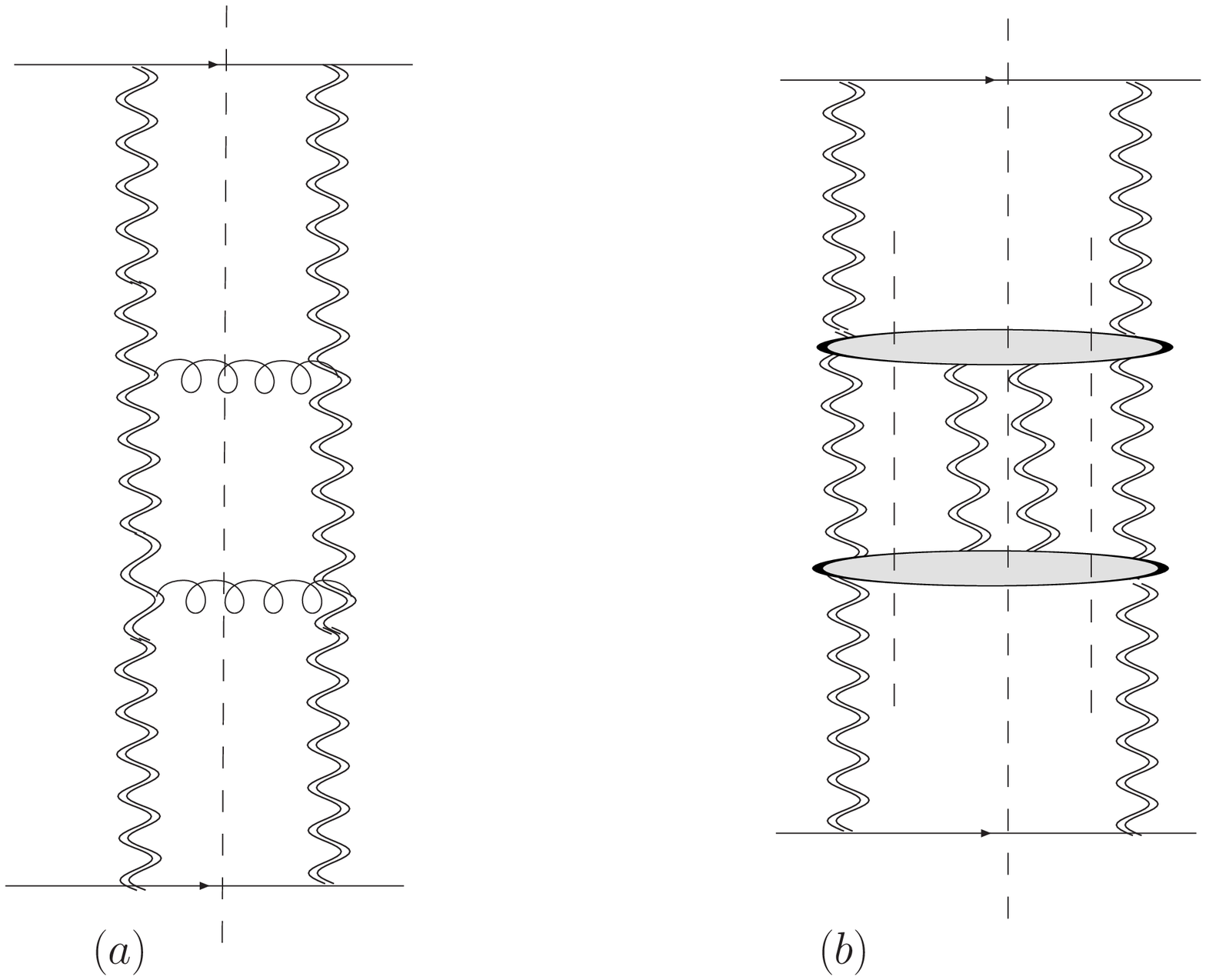,height=8cm}}
\caption{\em Quark quark scattering in the high energy 
limit of QCD (color singlet exchange): (a) two loop correction in the ladder approximation, 
(b) diagrams with two Triple Pomeron Vertices (grey bloobs).}
\label{fig:diagramy1}
\end{figure}
\begin{figure}[t!]
\centerline{\epsfig{file=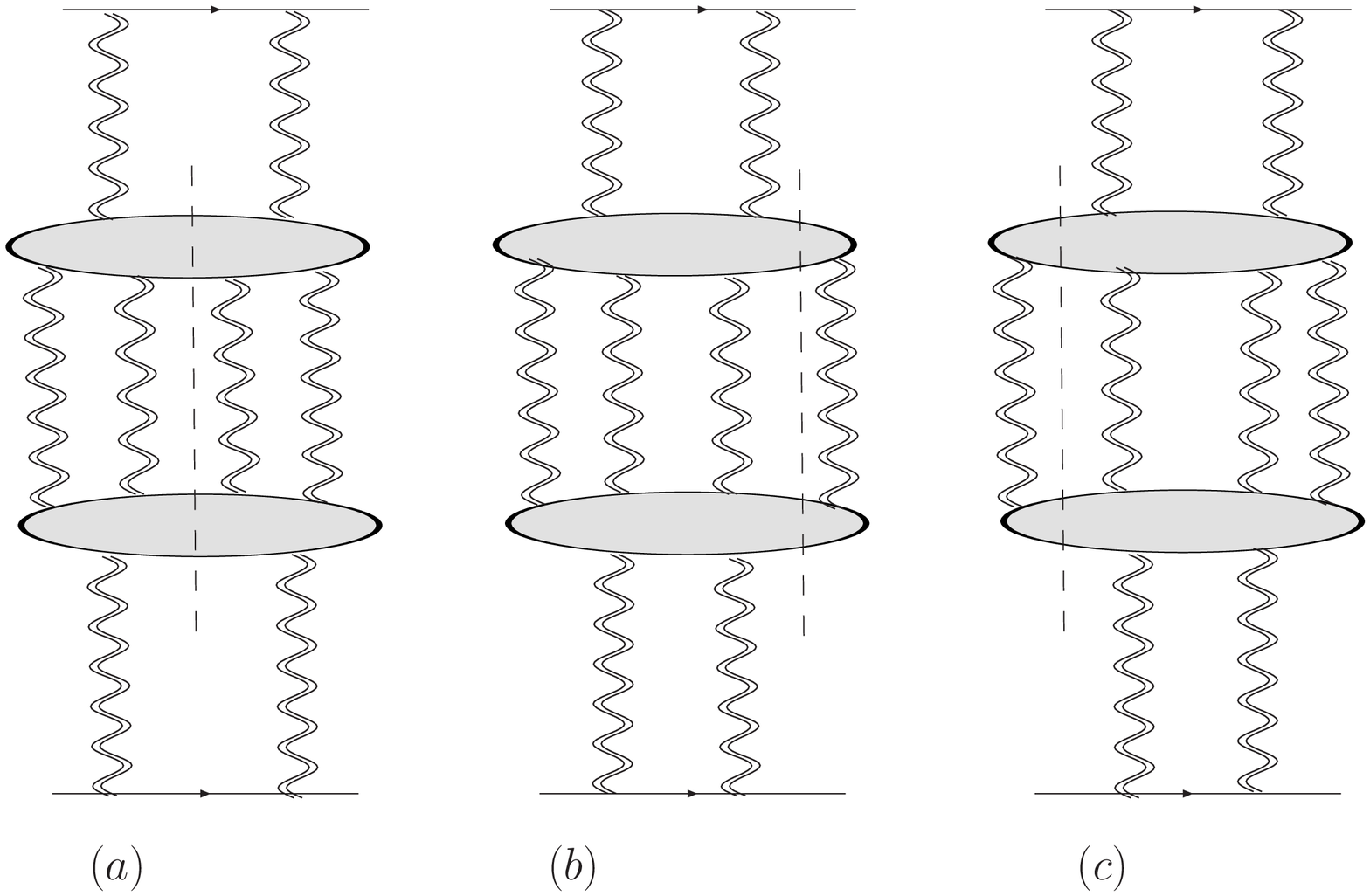,height=8cm}}
\caption{\em Different cuts.}
\label{fig:diagramy}
\end{figure}
Let us briefly explain how we obtain the correct normalization factors.
We consider the scattering amplitude of quark quark scattering 
(Fig.\ref{fig:diagramy1}b) 
which, in the center, contains the insertion of four reggeized $t$-channel  
gluons. We begin with the diffractive cut (Fig.\ref{fig:diagramy}a): the  
one loop amplitude on the lhs of the cutting line will be assumed to have an 
even signature exchange and, hence, is equal (up to factor $i$) to its energy 
disconinuity. As a result, we start from Fig. \ref{fig:diagramy1}b with 
three cutting lines: all horizontal lines are on mass-shell. To distribute the 
color and phase space factors we proceed as follows: using the on-shell 
conditions, we can perform $8$ out of the $10$ longitudinal Sudakov 
integrations; the two remaining longitudinal variables denote the rapidity of 
the two produced gluons in the central region which, for the moment, 
we keep fixed. For each closed loop, we are left with an integral 
$\int d^2k/(2\pi)^3$. For each BFKL rung in Fig.\ref{fig:diagramy1}a 
we have the kernel from (\ref{eq:2to2kernel}), for each $2 \to 4$ vertex 
in Fig.\ref{fig:diagramy1}b we have the $2 \to 4$ vertex (\ref{eq:fvertex}) 
(divided by the additional factor $1/\sqrt{2}$). 
Having in mind that our discussion should be applicable also to more 
general diagrams, we retain, for the moment, the general color structure 
in (\ref{eq:2to2kernel}) and in (\ref{eq:fvertex}). 
When deriving, from Fig.\ref{fig:diagramy1}b, the diffractive cut in 
Fig.\ref{fig:diagramy}a, we write, for each $2$ gluon exchange on the lhs 
and on the rhs of the cutting line a statistic factor $1/2!$; 
from the compensating factor $4$ we absorb $\sqrt{2}$ into each of the 
$2 \to 4$ vertices. As a result, we have, in addition to all other color and 
phase space factors, the statistic factors $2/(2!)^2$. Invoking now the 
AGK rules ~\cite{AGK,BVS}, applied to the exchange of four (odd signature) reggeized gluons, 
the other contributions in Figs.\ref{fig:diagramy}b and c, 
the statistic factors become:
\be
2 \left( \frac{1}{2!2!} - \frac{2}{3!}\right)  = - \frac{4}{4!}.
\ee  

Finally, we use our result for quark-quark scattering and return 
to the process of our interest, photon-photon scattering. 
Replacing the quark impact factors by photon impact factors, inserting 
BFKL rungs above and below the $2 \to 4$ vertices, and 
inserting pairwise interactions between the four gluon lines in the center,
we arrive at:
\bea
A(s,t)&=&2 i s\pi
\int_0^{Y} dY_3\int_0^Y dY_2\int_0^Y dY_1 \delta(Y-Y_1-Y_2-Y_3) \nonumber \\
&&\cdot \int \frac{d^2\kpp}{(2\pi)^3} \frac{d^2\kpp_1}{(2\pi)^3}
\phi^{a'_1a'_2}(\kpp,\q-\kpp){\cal G}^{(2)a_1'a_2';a_1''a_2''}(Y_3;\kpp,\kpp_1,\q)
\nonumber\\
&&\cdot \frac{-4}{4!}  \int \prod_{i=1}^4 \left( \frac{d^2k_i}{(2\pi)^3} \right) (2\pi)^3
\delta^{(2)}(\sum \k_i - \q)  \int \prod_{i=1}^4 \left(
\frac{d^2\k_i'}{(2\pi)^3}\right) (2\pi)^3  \delta^{(2)}(\sum \k_i' - \q)
\nonumber \\ 
&&\cdot {\cal V}^{a_1''a_2'';a_1a_2a_3a_4}(\kpp_1,\q-\kpp_1;\k_1,\k_2,\k_3,\k_4)
{\cal G}^{(4);\{a_i\}\{b_i\}}(Y_2;\{\k_i\} \{\k_i'\})
\nonumber \\
&&\cdot \int \frac{d^2\kpp'}{(2\pi)^3} \frac{d^2\kpp_1'}{(2\pi)^3}
{\cal V}^{b_1b_2b_3b_4;b_1''b_2''}(\k_1',\k_2',\k_3',\k_4';\kpp'_1,\q-\kpp'_1)
\nonumber\\
&&\cdot {\cal G}^{(2)b_1''b_2'',b_1'b_2'}(Y_1;\kpp'_1,\kpp',\q)
\phi^{b'_1b'_2}(\kpp',\q-\kpp'),
\label{eq:loop2}
\eea
where the minus sign in the third line indicates that the four gluon 
insertion into the two gluon Green's function represents a negative correction 
to the simple ladder amplitude.
Here $s$ is the squared center of mass energy, $Y\!\!=\!\ln(s/s_0)$ is the total
rapidity, $Y_1$, $Y_2$, $Y_3$ are the rapidity intervals as depicted 
in Fig. \ref{fig:pomloopBKP3}, $\phi^{a'_1a'_2}$ denotes the impact factor of the virtual 
photon, ${\cal G}^{(2)a_1'a_2';a_1''a_2''}_{\omega}$ 
is the BFKL Green's function which satisfies the BFKL integral equation: 
\bea
(\omega -\omega(\k_1)-\omega(\k_2))
{\cal G}^{(2)\;a_1 a_2 b_1 b_2}_{\omega} (\{\k_i\},\{\k'_i\})=&\nonumber\\
{\cal G}^{(2)0\;a_1 a_2 b_1 b_2}(\{\k_i\},\{\k'_i\})\;
+ \; \frac{1}{\k_1^2 \k_2^2}
K_{2\rightarrow 2}^{\{a\}\rightarrow\{b\}}\,\otimes\,
{\cal G}^{(2)\;a_1 a_2 b_1 b_2}_{\omega} (\{\k_i\}\{\k'_i\}),
&
\eea
with an inhomogeneous term analogous to (\ref{G40}). 
The statistics factor $\frac{1}{4!}$ reflects the symmetry of
the expression under the interchange of the four gluons. 
In eq.(\ref{eq:loop2}), the selfenergy is defined by lines 3 - 5, i.e. 
the convolution of the two $2 \to 4$ vertices with the BKP Green's function 
between them.
As a convenient simplification, we approximate the four gluon state by two noninteracting 
color singlet ladders (Fig. \ref{fig:pomloopBKP3}b): this configuration represents 
a 'pomeron loop'. 
It is easy to find the combinatorial factor of a system where two pairs of 
gluons form bound states. We have three possibilities of pairing two gluons to form bound states out of
four gluons. This yields the factor $1/2!$. In this configuration we have a pomeron loop topology.
The result reads:
\bea
A(s,t)=2 i s\pi 
\int_0^{Y} dY_3 \int_0^Y dY_2 \int_0^Y dY_1\;
\delta(Y-Y_1-Y_2-Y_3)\nonumber\\
\cdot \int\!\!\!\frac{d^2\kpp}{(2\pi)^3}\frac{d^2\kpp_1}{(2\pi)^3}
\phi^{a'_1a'_2}(\kpp,\q-\kpp)
{\cal G}^{(2)a_1'a_2',a_1^{''}a_2^{''}}(Y_3;\kpp,\kpp_1,\q)
\nonumber\\
\cdot \frac{-1}{2!} \int \frac{d^2\r}{(2 \pi)^3} \int \frac{d^2\k_1}{(2\pi)^3}
\frac{d^2\k_3}{(2\pi)^3}{\cal V}^{a_1''a_2'';a_1a_2a_3a_4}
(\kpp_1,\q-\kpp_1;\k_1,-\k_1-\r,\k_3,-\k_3+\r+\q)\nonumber\\
\cdot \int\frac{d^2\k_1'}{(2\pi)^3}\frac{d^2\k_3'}{(2\pi)^3}
(P{\cal G})^{(2)a_1a_2b_1b_2}(Y_2;\k_1,\k_1',\r)
(P{\cal G})^{(2)a_3a_4b_3b_4}(Y_2;\k_3,\k_3',\r+\q) \nonumber\\
\cdot \int\frac{d^2\kpp'_1}{(2\pi)^3}\frac{d^2\kpp'}{(2\pi)^3}
{\cal V}^{b_1b_2b_3b_4;b_1''b_2''}
(\k_1',-\k_1'-\r,\k_3',-\k_3'+\r+\q;\kpp_1',\q-\kpp_1')\nonumber\\
\cdot {\cal G}^{(2)b_1''b_2'',b'_1b'_2}(Y_1;\kpp_1',\kpp',\q)
\phi^{b'_1b'_2}(\kpp',\q-\kpp'),
\label{eq:loop1}
\eea
where 
\be
P^{a_1a_2b_1b_2}=\frac{\delta^{a_1a_2}\delta^{b_1b_2}}{N_c^2-1}
\ee
is the color singlet projector.
These projectors act on the color tensors of the $2 \to 4$ vertices, turning the pairs of color 
labels $(a_1 a_2)$, $(b_1 b_2)$, $(a_3 a_4)$, $(b_3 b_4)$ into color singlets.
Comparison with (\ref{eq:fvertex}) shows that this projection operator, 
when acting on the first term, leads to a factor $1$, whereas the remaining 
terms come with the weight factor $\frac{1}{N^2-1}$: in comparison with the first term, 
they are color suppressed.
This large-$N_c$ approximation turns the $2 \to 4$ vertices into the Triple Pomeron Vertices (TVP).  

\begin{figure}[t!] \centerline{\epsfig{file=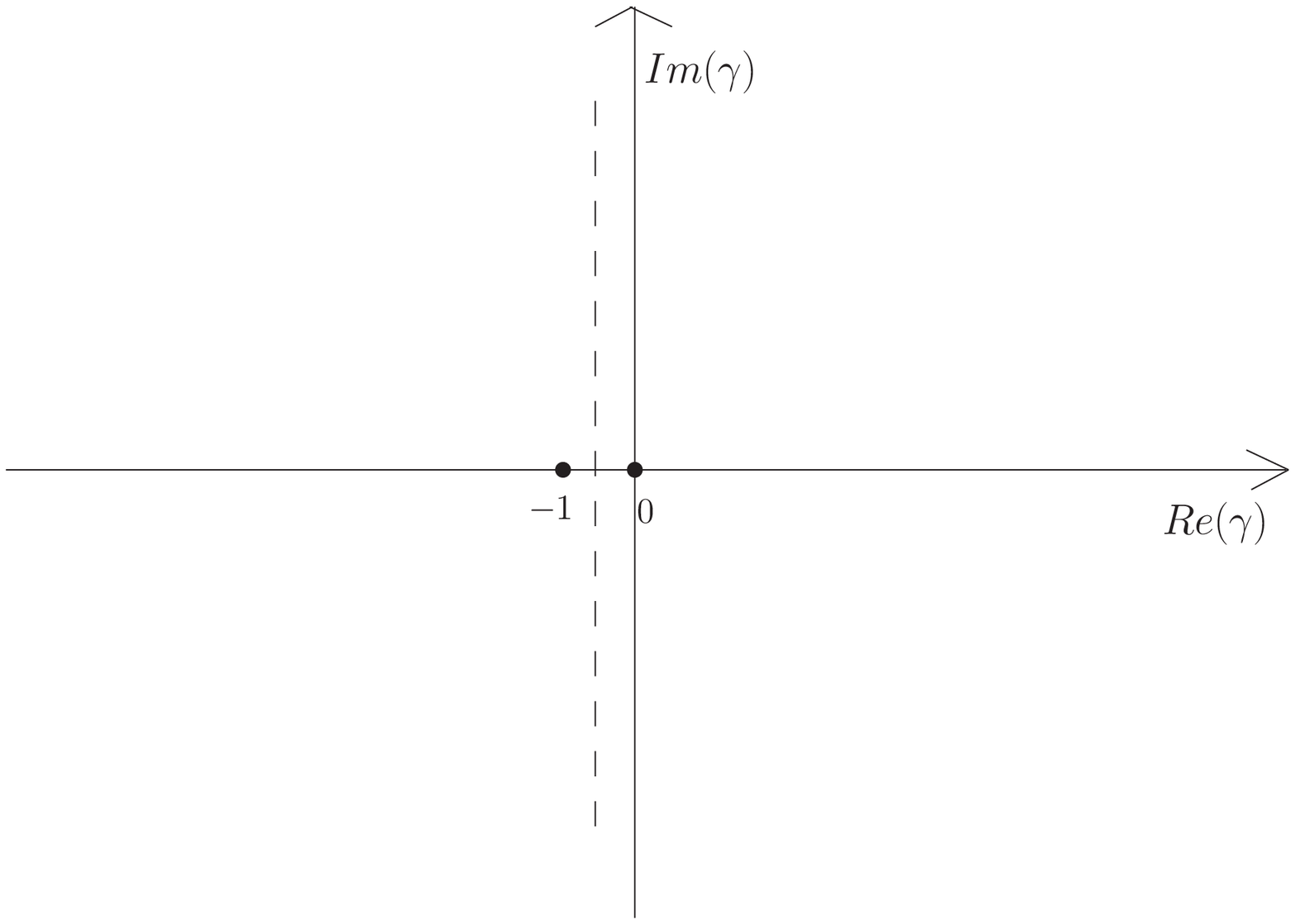,height=8cm}}
\caption{\em Singularities in the $\gamma$ plane.}
\label{fig:mellinfig}
\end{figure}

In the following we shall focus on the pomeron loop (\ref{eq:loop1}) and
investigate, for zero total momentum transfer,  
$\q=0$, the kinematic limit where the momentum scale of the upper
photon is much larger than the lower one. This implies that, at the upper
TPV, the momentum from above, $\kpp_1$, is larger than the momenta from
below, $\k_1$, $\k_3$, and the loop momentum $\r$ ('collinear limit'). 
Conversely, for the
lower TPV we have the opposite situation: the momenta $\k_1'$, $\k_3'$,
and $\r$ are larger than $\kpp_1'$ ('anticollinear limit').
Let us become a bit more formal. We expand the amplitude of Fig. \ref{fig:pomloopBKP3} 
in powers of $Q_0^2/Q_1^2$ ('twist expansion'). The object of our interest is the
self-energy of the Pomeron Green's function, $\Sigma(\kpp_1,\kpp'_1)$.
In Eq.(\ref{eq:loop1}), $\Sigma(\kpp_1,\kpp'_1)$ is defined to represent the 
lines 3 - 5, i.e. the convolution of the two TPV's with the two BFKL Green
functions between them. It has the dimension $\k^2$, and it is convenient to
define the dimensionless object $\tilde\Sigma(\frac{\kpp_1}{\kpp'_1})=
\frac{\Sigma(\kpp_1,\kpp'_1)}{\sqrt{\kpp_1^2\kpp_1^{'2}}}$
with the Mellin transform:
\be
\tilde\Sigma(\gamma)=\int_0^{\infty}dk^2\tilde\Sigma(k^2)(k^2)^{\gamma-1}.
\label{eq:mellin1}
\ee
The inverse Mellin transform reads:
\be
\tilde\Sigma(k^2)=\int_C \frac{d\gamma}{2\pi i}(k^2)^{-\gamma}
\tilde \Sigma(\gamma),
\label{mellintransform}
\ee
where $k^2=\frac{\kpp_1^{'2}}{\kpp_1^{2}}$, and the contour crosses the real axis
between $-1$ and $0$ (see Fig. \ref{fig:mellinfig}).
Our analysis will then reduce to the study of the singularities
of the function $\tilde\Sigma(\gamma)$. The twist expansion corresponds to
the analysis of the poles located to the left of the contour
in the $\gamma$ plane: the pole at $\gamma=-1$ is the leading twist pole,
the pole at $\gamma=-2$ belongs to twist 4, and so on.
As we have already said before, for the upper TPV in Fig. \ref{fig:pomloopBKP3}, 
the analysis of this twist expansion requires the 'collinear limit', for the lower TVP 
the 'anticollinear' one.

\section{The collinear limit}
In this section we are going to study the collinear limit of the TPV.
The ordering of the transverse momenta is the following:
$|\kpp_1|\equiv|\k|\!\!\gg\!\!|\k_1|,|\k_2|,|\k_3|,|\k_4|$. We therefore 
expand in powers of $|\k_1|/|\k|$, $|\k_2|/|\k|$, $|\k_3|/|\k|$,
$|\k_4|/|\k|$.
In our investigations we will be interested in attaching color singlet 
objects to the vertex, and we project (\ref{eq:fvertex}) onto the color 
singlets. In the limit $N_c \to \infty$ we obtain:
\bea
&P^{a_1 a_2 b_1 b_2} P^{a_3 a_4 b_3 b_4} 
{\cal V}^{a'_1 a'_2;a_1 a_2 a_3 a_4}(\k,-\k;\k_1,\k_2,\k_3,\k_4) = \nonumber \\
= & \delta^{a_1',a_2'} \delta^{b_1,b_2} \delta^{b_3,b_4} \frac{\sqrt{2}\pi}{N_c^2-1}
\bigg[ V(1,2,3,4) +\frac{1}{N^2-1} \left( V(1,3,2,4) + V(1,4,2,3)\right) \bigg],
\label{eq:kovl}
\eea
where $V(1,2,3,4)\equiv V(\k,-\k;\k_1,\k_2,\k_3,\k_4)$. The first term 
will be denoted by 
\be
{\cal V}_{L0N_c}^{\{a'\}\{b\}}(1,2,3,4)= \delta^{a_1',a_2'} \delta^{b_1,b_2} \delta^{b_3,b_4} 
\frac{\sqrt{2}\pi}{N_c^2-1} V(1,2,3,4),
\ee
the second and third ones by 
\be
{\cal V}_{subN_c}^{\{a'\}\{b\}}(1,3,2,4)= \delta^{a_1',a_2'} \delta^{b_1,b_2} \delta^{b_3,b_4} 
\frac{\sqrt{2}\pi}{(N_c^2-1)^2} V(1,3,2,4)
\ee
etc. 
\subsection{The real part}
Let us begin the analysis by expanding the real part of the $G$ function (\ref{eq:funkg1}) in the
collinear limit. As we are going to limit ourselves to
the forward case we use the
simplified notations $G(\k,-\k;\k_1,-\k_1-\k_3,\k_3)\equiv G(\k_1,\k_3)$
and $V(\k,-\k;\k_1,\k_2,\k_3,\k_4)\!\equiv\!V(\k_1,\k_2,\k_3,\k_4)$.
With these definitions the $G_1$ function (\ref{eq:funkg2}) reads:
\be
\begin{split}
G_1(\k_1,\k_3)\,=\,
\frac{\k_1^2\k^2}{(\k-\k_1)^2}+
\frac{\k_3^2\k^2}{(\k+\k_3)^2}-\frac{(\k_1+\k_3)^2
\k^4}{(\k-\k_1)^2(\k+\k_3)^2}. 
\label{eq:greal}
\end{split}
\ee
In the collinear limit the momenta of the outgoing gluons satisfy the 
conditions 
$|\k_1|\!\!<\!<\!\!|\k|$, $|\k_3|\!\!<\!<\!\!|\k|$. Performing the expansion 
in $|\k_i|/|\k|$ up to fourth order terms we obtain:
\be
\begin{split}
G_1(\k_1,\k_3) = 2\k^2\Bigg[-\frac{\k_1\cdott\k_3}{\k^2}
-\frac{\k_1\cdott \k}{\k^4} 
\left( \k_3^2 + 2 \k_1 \k_3 \right) 
+\frac{\k_3\cdott \k}{\k^4} 
\left(\k_1^2  + 2 \k_1 \k_3 \right) + \\
+ \left( \frac{\k_1^2}{\k^2} - (\frac{2\k_1 \cdott \k}{\k^2})^2 \right)   
                \left(\k_3^2 + 2 \k_1 \k_3 \right) 
+ \left( \frac{\k_3^2}{\k^2} - (\frac{2\k_3\cdott \k}{\k^2})^2 \right)
                 \left(\k_1^2 + 2 \k_1 \k_3 \right) + \\
+ (\k_1+\k_3)^2 \frac{2\k \cdott \k_1}{\k^2} \frac{2\k \cdott \k_3}{\k^2}
           +...\Bigg].
\label{eq:G1}
\end{split}
\ee
The first term is the twist-two contribution:
\be
G_{1}(\k_1,\k_3)^{\tau=2}\, =\,-2\k^2\,\frac{\k_1\cdott \k_3}{\k^2}.
\ee
With analogous expressions for the other $G_1$ functions in eq.(\ref{eq:vertex}) 
we find that the sum of all twist-two pieces vanishes. The next two terms 
on the rhs of eq.(\ref{eq:G1})
vanish after averaging over the azimuthal angle of $\k$.
Finally we are left with the twist-four piece. After averaging over the 
direction of $\k$ we find:
\begin{equation}
G_{1}(\k_1,\k_3)^{\tau=4}\,=\,2\k^2\left[\frac{2 (\k_1\cdott \k_3)^2-\k_1^2
\k_3^2}{\k^4}\right].
\end{equation}
From eq.(\ref{eq:vertex}) we obtain for the twist-4 piece of the real part 
of the TPV:
\be
\cVr(\k_1,\k_2,\k_3,\k_4)^{\tau=4}\!=\nonumber 
\ee
\be
\delta^{a_1',a_2'} \delta^{b_1,b_2} \delta^{b_3,b_4} 
\frac{\sqrt{2}\pi}{N_c^2-1} \frac{g^4}{2} 2\k^2
\frac{4(-\k_1\cdott \k_2\,\k_3\cdott \k_4+
\k_1\cdott \k_4\,\k_2\cdott \k_3+\k_1\cdott \k_3\,\k_2\cdott \k_4)}{\k^4},
\ee
where the superscript $r$ stands for the real emission. This expression is the master formula 
for the twist-four contribution.

In the next step we are going to attach BFKL ladders to the pairs of gluons $(\k_1,\k_2)$ and 
$(\k_3,\k_4)$. Since the presence of a momentum transfer across the BFKL ladder would cause loss of a
logarithmic contribution, we limit ourselves to the forward directions: 
\be
\k_1=-\k_2,\,\,\,\,\k_3=-\k_4.
\ee
Putting $\k_1=\l$, $\k_2=-\l$, $\k_3=\m$, $\k_4=-\m$ we obtain:
\begin{equation}
\cVr(\l,-\l,\m,-\m)^{\tau=4}\,=\,\delta^{a_1',a_2'} \delta^{b_1,b_2} 
\delta^{b_3,b_4} 
\frac{\sqrt{2}\pi}{N_c^2-1}
\frac{g^4}{2}2\k^2\frac{4(2(\l\cdott \m)^2 - \l^2 \m^2)}{\k^4}.
\end{equation}
Now we multiply the vertex by propagators for the lower gluon lines and convolute with 
$2\to2$ transition kernels (eq. (\ref{eq:2to2kernel})). 
Our goal is to find, from the convolutions of the vertex with propagators and kernels, the maximal 
number of logarithms. To do that we should act on the twist four contribution of the vertex 
with the twist four evolution operator, which, in our case, is the product of two BFKL kernels 
in the twist-two approximation.
\begin{figure}[t!]
\centerline{\epsfig{file=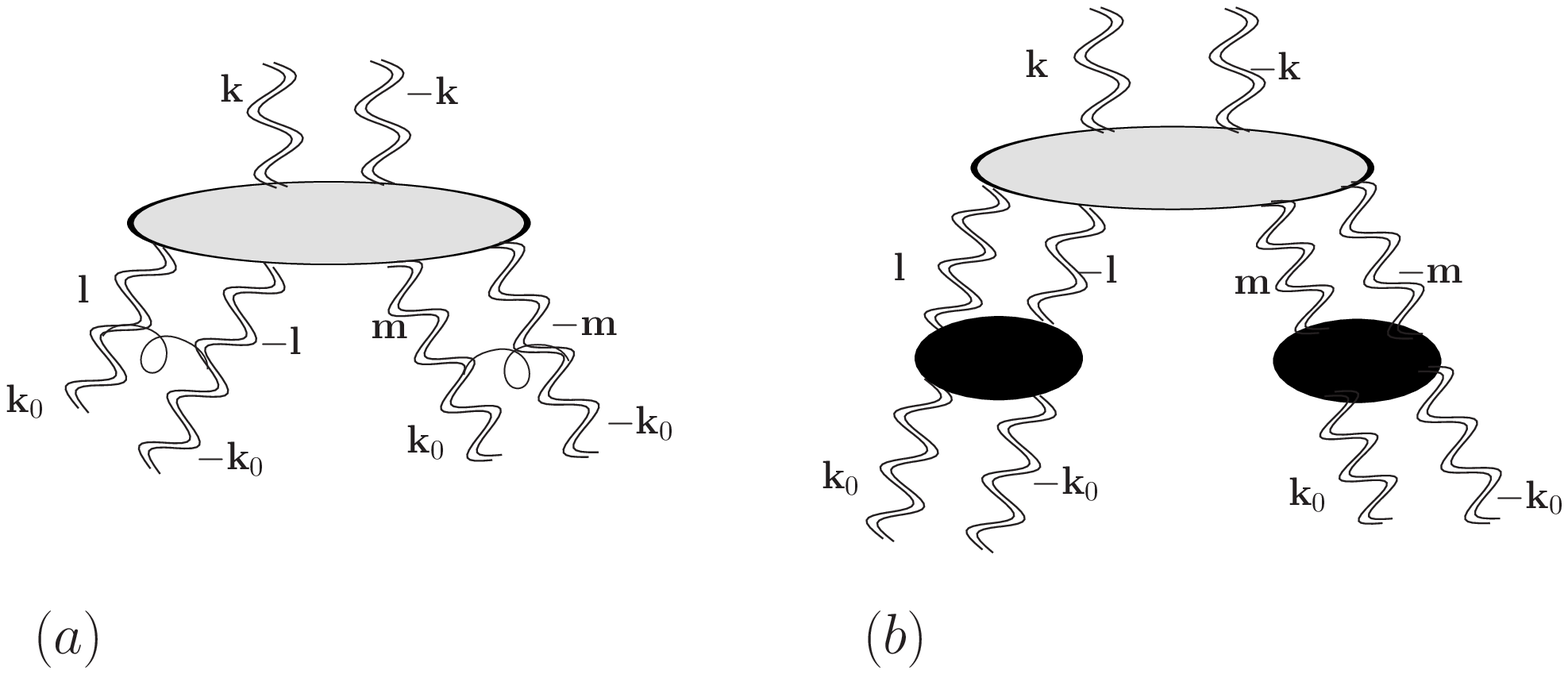,height=4.5cm}}
\caption{\em (a) The TPV with two BFKL interactions attached to it.
(b) The TPV with gluon ladders.}
\label{fig:drabiny}
\end{figure}
Let us compute  the collinear approximation to the BFKL kernels which, when
convoluted with the TPV, will give a logarithmic integral.
The expression for the emission part of the BFKL kernel is: 
\be
K(\q_1,\q_2;\k_1,\k_2)= - N_cg^2\bigg[(\k_1+\k_2)^2 -\frac{\q_2^2
\k_1^2}{(\k_2-\q_2)^2}-\frac{\q_1^2 \k_2^2}{(\k_1-\q_1)^2}\bigg].
\label{eq:eqBFKL}
\ee
The factor $-N_c$ replaces the color tensors in eq.(\ref{eq:2to2kernel}), 
since we have projected on the color singlet state.
Assuming zero momentum transfer and $\q^2_1\!>\!\!\!>\!\k^2_2$ we get:
\be
K\!=g^2N_c2\k_1^2,
\ee
where, in order to simplify the notation, we have skipped the arguments of the kernel $K$.
Using this approximation in the formula for Fig.\ref{fig:drabiny}a, we get the following expression for 
the convolution of the vertex with two BFKL kernels (one kernel for each 
two-gluon pair below the vertex):
\bea
(K\,K)\otimes
{\cVr}^{\tau=4}&\,=\,& \delta^{a_1',a_2'} \delta^{b_1,b_2} \delta^{b_3,b_4} 
\frac{\sqrt{2}\pi}{N_c^2-1} N_c^2 2\k^2 \frac{g^8}{2}\cdot\nonumber\\
&&\int_{\k_0^2}^{\k^2}\frac{d^2\l}{(2\pi)^3}\int_{\k_0^2}^{\k^2}\frac{d^2\m}{(2\pi)^3}
\frac{2\k_0^2}{\l^4}\frac{2\k_0^2}{\m^4}
\frac{4(2(\l\cdott \m)^2 - \l^2 \m^2)}{\k^4}\nonumber\\& =&0
\label{eq:lmvet},
\eea
where $\k_0$ is the lowest momentum scale which we do not specify at present.
We notice that the integrals of $\l$ and $\m$ are logarithmic, but - what is 
most striking - the angular integral over the angle between $\l$ and $\m$ 
renders the Triple Pomeron Vertex to vanish. It is straightforward to iterate the convolution with 
BFKL kernels (Fig. \ref{fig:drabiny}b), and, as a result, we arrive at the conclusion 
that - after averaging over the azimuthal angles - the twist-four part of 
the TPV vertex gives zero contribution.

\subsection{The virtual part} 
So far we have investigated contributions coming from the real part of the 
vertex. What remains are the disconnected parts. 
In order to investigate logarithmic contributions of the virtual pieces 
we have to convolute them with an impact factor at the upper end of 
Fig.\ref{fig:diagramy2}b. To deal with infrared finite quantities 
it is convenient to work with the impact factor of the photon. 
The function $G_2(\k_1,\k_2+\k_4,\k_3) = G_2(\k_1,\k_3)$ (\ref{eq:funkg3}) in the forward direction 
reads: 
\be
G_2(\k_1,\k_3)=-\k^4 \frac{1}{8\pi^2} 
\left(\ln\frac{|\k_1|^2}{|\k_1+\k_3|^2}\delta^{(2)} (\k_1-\k)
+ \ln\frac{|\k_3|^2}{|\k_1+\k_3|^2}\delta^{(2)} (\k_3+\k) \right).
\label{eq:gvirt}
\ee
The photon impact factor (for transversely polarized photons) has the form 
~\cite{Nik,AM}
\be
\phi_{a_1'a_2'}(\k,Q)=\delta^{a_1'a_2'}\alpha_s\alpha_{em}\sum_q e_q^2\int_0^1 d\tau
d\rho\frac{[\rho^2-(1-\rho)^2]
[\tau^2-(1-\tau)^2]\k^2}{\rho(1-\rho)Q^2+\tau(1-\tau)\k^2},
\ee
where $Q^2$ and $\k^2$ denote the (negative) photon and gluon virtualities, 
resp.
In this expression $\rho$ denotes the longitudinal component of the quark loop 
momentum (in the Sudakov decomposition), while the 
second integration variable $\tau$ is a Feynman parameter.
For our investigations we are interested in the twist expansion.
To perform the twist expansion of the impact factor one has to perform the 
Mellin transform with respect to $\k^2/Q^2$.
With the Mellin transform (\ref{eq:mellin1}) one finds:
\be
\phi_{a_1'a_2'}(\k,Q)=\int\frac{d\gamma}{2\pi
i}\left(\frac{\k^2}{Q^2}\right)^{-\gamma}\phi_{a_1'a_2'}(\gamma),
\label{fig:impactmellin}
\ee
and we obtain:
\be
\phi_{a_1'a_2'}(\gamma)=\delta^{a_1'a_2'}{\cal C}\frac{\Gamma(3+\gamma)}
{\frac{5}{2}+\gamma}\frac{\Gamma(1+\gamma)}{1+\gamma}
\frac{\Gamma(-\gamma)}{-\gamma}\frac{\Gamma(-\gamma+2)}{\Gamma(-\gamma+
\frac{3}{2})}.
\ee
Turning to Fig.\ref{fig:pomloopBKP3}b, we are interested in the following 
ordering of momenta: $|Q_1|\!\gg\!|\kpp|\!\gg\!|\kpp'|\!\gg\!|Q_0|$.
To analyse the twist-4 term of this kinematic region we need, for the upper 
impact factor, the twist-4 term. Closing, in eq.(\ref{fig:impactmellin}), the contour of the 
$\gamma$ integration to the left we obtain the following collinear expansion 
of the photon impact factor:
\bea
\phi_{a_1'a_2'}(\kpp,Q_1)=\delta^{a_1'a_2'}\phi(\kpp,Q_1)\,=\,
\delta^{a_1'a_2'}\B\Bigg\{\left[\frac{14}{9}-\frac{4}{3}\ln\left(\frac{\kpp^2}{Q_1^2}\right)\right]
\frac{\kpp^2}{Q_1^2} +
\frac{2}{5}\left(\frac{\kpp^2}{Q_1^2}\right)^2+...\Bigg\}\,,
\label{eq:imp}
\eea
where $\B=\sum_f e_f^2\alpha_s\alpha_{em}$.
As it is well-known, the twist-four term has no logarithmic enhancement.  

For later purposes we also list the results for the lower impact factor: we close the contour to the right and find:
\be
\phi_{a_1'a_2'}(\kpp',Q_0)=\delta^{a_1'a_2'}\phi(\kpp',Q_0)\,=\,
\delta^{a_1'a_2'} 
\B\Bigg\{\left[\frac{14}{9}-\frac{4}{3}\ln\left(\frac{Q_0^2}{{\kpp'}^2}\right)
\right]+\frac{2}{5}\left(\frac{Q_0^2}{{\kpp'}^2}\right)^...\Bigg\}.
\label{impactanticollinear}
\ee
For our twist-4 analysis of Fig.\ref{fig:pomloopBKP3}b we will need 
the second term on the rhs.

Returning to the upper impact factor and concentrating on the twist-4 piece,
we now easily see, by simply counting powers of momenta, that the virtual 
contributions of the TPV cannot contribute to the maximal power of logarithms. 
Namely, beginning with the impact factor above the TVP,
we have the power $\k^4$ which cancels the two gluon propagators attached 
to the impact factor. From the $G_2$ functions we find another power, $\k^4$, 
which, through the $\delta$-functions, turns into $\m^4$, $\l^4$, or 
$(\m \pm \l)^4$. When dividing the region of integration into the two 
parts $\m \ll \l$ and $\l \ll \m$, the terms with $(\m \pm \l)^4$ 
turn into $\l^4$ or $\m^4$.  
Below the TPV we have the pairs of propagators, $1/\m^4$, and $1/\l^4$, 
and there is no $\m$ (or $\l$)- dependent contribution from the BFKL kernels.
Combining these momentum factors, we therefore obtain only 
integrals of the form 
$$ \int \frac{d^2 l}{\l^4} \int \frac{d^2m}{\m^4} \l^4$$
or
$$ \int \frac{d^2 l}{\l^4} \int \frac{d^2m}{\m^4} \m^4,$$
i.e.none of the integrals is logarithmic (this argument remains unaffected if 
we include the logarithms from the $G_2$ functions).    
Hence, within the leading-log approximation, also the virtual part of the 
TPV is zero.  

Let us emphasize that our search for the 'maximal power of transverse logs'
is exactly what is required for a consistent twist-four analysis. 
In order to obtain 
this maximal power (i.e. one power for each transverse momentum loop integral),
we had to start, at the upper impact factor with the twist-four term.
Including a BFKL-ladder between the impact factor and the TPV forces us 
to take also the twist-four approximation of the kernel, i.e. instead of the 
leading twist approximation in (\ref{eq:eqBFKL}), terms of the order
${\cal O}(\k_1^2 \frac{\k_1^2}{\q_1^2})$. Next, at the TPV we searched for terms 
of the order $\frac{\m^2 \l^2}{\k^2}$, and, finally, for the two BFKL kernels 
below the TPV, again the twist-2 approximation (\ref{eq:eqBFKL}). 
It is only this sequence of 
approximations which provides one logarithm for each loop, i.e. otherwise we 
loose one (or more) powers of logarithmic enhancements.   
Our result then says that one coefficient in this sequence of terms, 
namely the TPV, vanishes and thus makes the twist-four term in the twist 
expansion (in the leading-log approximation) disappear.

\subsection{Generalization to all higher twists}

The main result of the previous subsections -  the absence of collinear logarithms 
in the case of angular averaged BFKL ladders - 
can be generalized to all orders of powers of $1/Q^2$. We return to the function 
$V$ in (\ref{eq:vertex}) which is expressed in terms of the functions $G_1$ and 
$G_2$, and we average over the angles of $\m$ and $\l$. First $G_1$:
\be
G_1(\l,\m)=\frac{\k^2\l^2}{(\k-\l)^2}+\frac{\k^2\m^2}{(\k+\m)^2}
-\frac{\k^4(\l+\m)^2}{(\k-\l)^2(\k+\m)^2}.
\ee
Let us denote the first term in this formula by A, the second one by B and the 
third one by C, the angle between $\l$ and $\m$  by $\alpha$, and the angle between 
$\m$ and $\k$ by $\beta$ (the angle between $\l$ and $\k$ then equals 
$2 \pi-\alpha\!-\!\beta$). For the integrals over $\alpha$ and $\beta$ we find:
\be
I_A=\frac{1}{(2\pi)^2}\int_0^{2\pi}\!\!d{\alpha}\int_0^{2\pi}\!\!d{\beta}\,A=
\frac{\k^2\l^2}{|\l^2-\k^2|}
\ee
and:
\be
I_B=\frac{1}{(2\pi)^2}\int_0^{2\pi}d{\alpha}\int_0^{2\pi}d{\beta}\,
B=\frac{\k^2\m^2}{|\m^2-\k^2|}.
\ee
To compute the integral over $C$ we split $C=C_1+C_2+C_3$ into three pieces.
The corresponding integrals are:  
\be
I_{C_1}=\frac{1}{(2\pi)^2}\int_0^{2\pi}\!\!d\alpha d\beta \frac{\k^4 \l^2}{(\k^2-2\l\k\cos(\alpha+\beta)+\l^2)(\k^2+
2\m\k\cos\beta+\m^2)},
\ee
\be
I_{C_2}=\frac{1}{(2\pi)^2}\int_0^{2\pi}\!\!d\alpha d\beta
\frac{2 \k^4 |\l| |\m| \cos\alpha}{(\k^2-2\l\k\cos(\alpha+\beta)+\l^2)(\k^2+
2\m\k\cos\beta+\m^2)},
\label{eq:integral2}
\ee
\be
I_{C_3}=\frac{1}{(2\pi)^2}\int_0^{2\pi}d\alpha d\beta \frac{\k^4 \m^2}{(\k^2-2\l\k\cos(\alpha+\beta)+\l^2)(\k^2+
2\m\k\cos\beta+\m^2)}.
\ee
The results of the integration are:
\be
I_{C_1}=\frac{\l^2\k^4}{|\l^2-\k^2||\m^2-\k^2|},
\ee
\be
I_{C_2}=\frac{- 8\l^2\m^2\k^6}{|\l^2-\k^2||\m^2-\k^2|(\l^2+\k^2
+|\l^2-\k^2|)(\m^2+\k^2+
|\m^2-\k^2|)},
\ee
\be
I_{C_3}=\frac{\m^2\k^4}{|\l^2-\k^2||\m^2-\k^2|}.
\ee
The total contribution is given by summing up $I_A$, $I_B$, $I_{C_1}$, $I_{C_2}$ , $I_{C_3}$
The result can greatly be simplified if we consider special situations.
For instance, if $\k^2\!\ll\l^2,\m^2$, we may drop the absolute value signs. 
Adding all terms we obtain:
\be
\sum_{A,..,C_{3}} I=\frac{2\l^2\m^2\k^2-2\l^2\k^4-2\b^2\k^4+2\k^6}
{(\l^2-\k^2)(\m^2-\k^2)} =2 \k^2. 
\ee   
In all other cases: $\k^2\!\gg\l^2,\m^2$, $\m^2\!\!\gg\!\!\k^2\!\!\gg\!\!\l^2$, or
$\l^2\!\!\gg\k^2\!\!\gg\!\!\m^2$ 
the sum of all terms gives zero.
Therefore the final result can be simply written as:
\be
\frac{1}{(2\pi)^2}\int_0^{2\pi}\!\!d\alpha d\beta G_1(\l,\m)
=2\k^2\theta(\l^2-\k^2)\theta(\m^2-\k^2),
\ee
where the factor $1/(2\pi)^2$ comes from averaging.

Let us now perform the angular averaging of the disconnected pieces 
of the $G(\l,\m)$ function. We have:
\be
G_2(\l,\m)=-\k^4 \frac{1}{8\pi^2} \left( \ln\frac{\l^2}{(\l+\m)^2}
\delta^{(2)}(\l-\k) + \ln\frac{\m^2}{(\l+\m)^2}\delta^{(2)}(\m-\k)\right).
\ee 
To compute the integral over angles we split the region of integration.
In the case when $|\m|\!\!\gg\!\!|\l|$ the first term gives 
\be
I_D=\frac{1}{(2\pi)^2}\int_0^{2\pi}d\alpha d\beta 
\ln \frac{\l^2}{\l^2+\m^2+2\l \m \cos\alpha} = \ln\frac{\l^2}{\m^2},
\ee 
whereas the second one vanishes. 
In the case when $|\l|\!\!\gg\!\!|\m|$ we obtain zero from the first term and
\be
 I_D= \frac{1}{(2\pi)^2}\int_0^{2\pi}d\alpha d\beta 
\ln \frac{\m^2}{\l^2+\m^2+2\l \m \cos\alpha} = \ln\frac{\m^2}{\l^2}
\ee
from the second one. We combine the two cases in the following way: 
\be
G_2(\l,\m)=-\k^4 \frac{1}{8\pi^2} \left( \ln\frac{\l^2}{\m^2}
\theta(\m^2-\l^2)\delta^{(2)}(\l-\k)
+ \ln\frac{\m^2}{\l^2}\theta(\l^2-\m^2)\delta^{(2)}(\m-\k)\right).
\ee
 
Putting all pieces together and including the remaining $G$ functions 
we arrive at the angular-averaged form of $V$:
\bea
\frac{1}{(2\pi)^2}\int_0^{2\pi}d\alpha d\beta V(\k,-\k;\l,-\l,\m,-\m)=
4\frac{g^4}{2}\bigg[2\k^2\theta(\l^2-\k^2)\theta(\m^2-\k^2)\nonumber\\
+\frac{1}{8\pi^2}\bigg(\ln\left(\frac{\l^2}{\m^2}\right)
\delta^{(2)}(\l-\k) \theta(\m^2-\l^2)+
\ln\left(\frac{\m^2}{\l^2}\right)\delta^{(2)}(\m-\k)\theta(\l^2-\m^2)
\bigg)\bigg].
\label{eq:melG2}
\eea

The presence of the $\theta$-functions forbids all collinear configurations, i.e. there is no 
expansion in inverse powers of $\k^2$. The physical meaning of this result is
the following: if the two pomerons entering the vertex from below have smaller momenta than the
pomeron from above, they cannot resolve it and cannot merge because they do
not feel 'color' and the vertex vanishes\footnote {A similar result has first 
been noticed in ~\cite{GLR}: however, the disconnected pieces have been missed.
The result (\ref{eq:melG2}) agrees with the form given in ~\cite{BW}.}.
In the language of a twist-expansion, our result states that, in the 
leading-log approximation, not only twist-four, but all higher twist terms 
are zero, provided we restrict ourselves to the large-$N_c$ limit, and we 
use only the BFKL ladders with conformal spin zero below the TPV.     
  
\section{The anticollinear limit}
\subsection{Real part}
\begin{figure}[t!] \centerline{\epsfig{file=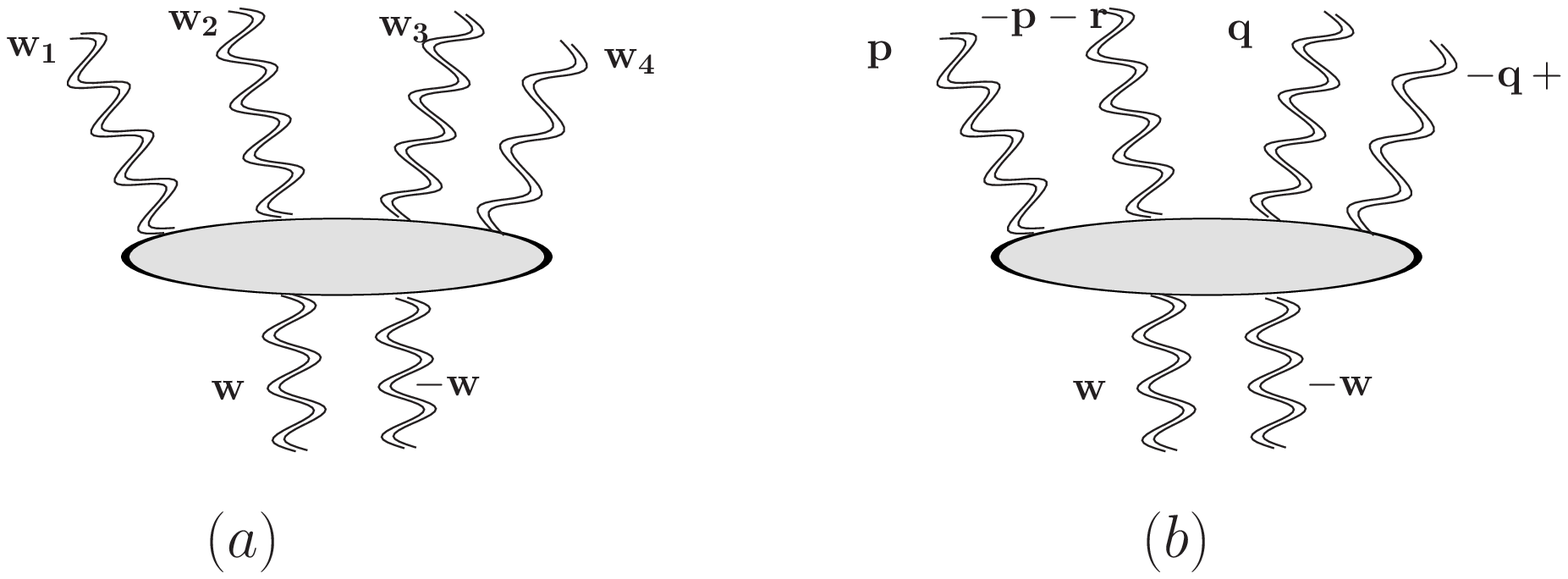,height=4.5cm}}
\caption{\em Momentum assignments at the lower TPV.}
\label{fig:TPVrealanticol}
\end{figure}

Let us now
investigate the anticollinear limit of the $2 \to 4$ vertex (Fig.\ref{fig:TPVrealanticol}).
In contrast to the collinear limit where the BFKL ladders below the vertex had to be 
in the forward direction, the anticollinear configuration allows for a nonzero momentum
transfer across the BFKL ladders above the vertex. 
The momentum transfer here, as we will see, does not lead
to a loss of a logarithm. We are interested in the limit 
$|\w|\!\!<\!<\!\!|\w_1|\!,|\w_2|\!,|\w_3|\!,|\w_4|$. 
To study the real emission part of the TPV it is convenient to rewrite the
$G_1$ function in the form: 
\bea
G_1(\w_1,\w_2+\w_3,\w_4)= \w^2\left[
\frac{1}{(1-2 \frac{\w\cdot\w_1}{\w_1^2}+ \frac{\w^2}{\w_1^2})}+
\frac{1}{(1+2 \frac{\w\cdot\w_4}{\w_4^2}+\frac{\w^2}{\w_4^2})} \right.\nonumber \\ \left.
- \w^2 \frac{(\w_1+\w_4)^2}{\w_1^2 \w_4^2} 
\frac{1}{(1-2 \frac{\w\cdot\w_1}{\w_1^2}+ \frac{\w^2}{\w_1^2})
 (1+2 \frac{\w\cdot\w_4}{\w_4^2}+\frac{\w^2}{\w_4^2})}\right].
\label{eq:Ganticol}
\eea
Here we have used the momentum conservation $\sum_i \w_i =0$.  The expansion parameters are $|\w|/|\w_1|$ and $|\w|/|\w_3|$. Performing
the expansion we obtain up to second order:
\bea
G_1(\w_1,\w_2+\w_3,\w_4)= \w^2\left[2+ 2 \frac{\w\cdott \w_1}{\w_1^2} - 2 \frac{\w\cdott
\w_4}{\w_4^2}  \right. \nonumber \\ \left.
- 2 \frac{\w^2}{\w_1^2} - 2 \frac{\w^2}{\w_4^2} - 2 \w^2 \frac{\w_1\cdott \w_4 }{\w_1^2 \w_4^2}  
+ \left( 2 \frac{\w \cdott \w_1}{\w_1^2} \right)^2 + \left( 2 \frac{\w \cdott \w_4}{\w_4^2} \right)^2 + ...\right].
\label{eq:rozwG}
\eea
Using (\ref{eq:funkg2}) we obtain for the leading term of the TPV:
\be
\cVrr(\p,-\p-\r,\q,-\q+\r)^{leading}=\delta^{a_1',a_2'} \delta^{b_1,b_2} 
\delta^{b_3,b_4} 
\frac{\sqrt{2}\pi}{N_c^2-1} \frac{g^4}{2} 2\w^2.
\ee
One easily sees that this term does not provide logarithms in the momentum scale. 
The subsequent terms in (\ref{eq:rozwG}) vanish after 
averaging over the angle of $\w$. Therefore, in order to get, after convolution with BFKL kernels in the subsystems 
$(12)$ and $(34)$, the required logarithmic contribution we need 
to consider, in (\ref{eq:rozwG}), terms of higher order. After averaging over the angle of $\w$, and after summing, in (\ref{eq:funkg2}), over all the 
$G_1$ functions, the resulting contribution is the following: 
\be
\cVrr(\w_1,\w_2,\w_3,\w_4)^{\tau=-2}=\nonumber 
\ee 
\be
\begin{split}
\delta^{a_1',a_2'} \delta^{b_1,b_2} \delta^{b_3,b_4} 
\frac{\sqrt{2}\pi}{N_c^2-1} g^4 \w^4    
               \Bigg[-\frac{\w_1\cdott \w_3}{\w_1^2
\w_3^2}-\frac{\w_2\cdott\w_3}{\w_2^2 \w_3^2} -\frac{\w_1\cdott \w_4}{\w_1^2
\w_4^2}-\frac{\w_2\cdott \w_4}{\w_2^2 \w_4^2} \\
-\frac{\w_1\cdott (\w_1+\w_2)}{\w_1^2 (\w_1+\w_2)^2}-
                     \frac{\w_2\cdott (\w_1+\w_2)}{\w_2^2 (\w_1+\w_2)^2}-
                     \frac{\w_3\cdott (\w_3+\w_4)}{\w_3^2 (\w_3+\w_4)^2}-
                     \frac{\w_4\cdott (\w_3+\w_4)}{\w_4^2 (\w_3+\w_4)^2}\\
                   -\frac{(\w_1+\w_2)\cdott(\w_3+\w_4)}
                      {(\w_1+\w_2)^2(\w_3+\w_4)^2}\Bigg].
\end{split}
\label{eq:mastac}
\ee

To proceed further we need the
anticollinear limit of the BFKL kernel. Using (\ref{eq:eqBFKL}), setting
$\k_1=\w_1$, $\k_2=\w_2$ and
requiring that $|\q_1|\!\!\gg\!\!|\w_1|\!,|\w_2|$ we obtain:
\be
K=-\, g^2N_c2\w_1\cdott\w_2.
\label{eq:approxkern}
\ee
As already mentioned before we are interested in the maximal power of 
logarithms in the momentum 
scale; this leads to the particular momentum configuration, where
the momentum transfer across the BFKL Pomerons in the subsystems 
$(12)$ and $(34)$ is nonzero.
(note that below the vertex we are still in the forward direction).
We set:
$\w_1\!\!=\!\p$, $\w_2\!\!=\!-\p\!-\!\r$,$\w_3\!=\!\q$,
$\w_4\!=\!-\q\!+\!\r$.
In order to obtain, after convoluting with BFKL kernels in the subsystems $(12)$ and $(34)$, logarithmic contributions,
we have to consider the following momentum-ordered configurations:
\begin{itemize}
\item The configuration where $|\r|\!\!\gg\!|\p|\!\gg\!|\q|$. 
BFKL kernels and propagators are of the form: 
\be
N_cg^2 2\p\cdott \r,\,\,\, \,\, - N_cg^2 2\q\cdott \r
\label{BFKLcoll}
\ee
and 
\be
\frac{1}{\p^2\r^2}\,\, ,\,\, \frac{1}{\q^2\r^2}, 
\label{propagators}
\ee
resp.
In order to render all transverse momemtum integrations (in $\p$, $\q$, and in $\r$) logarithmic, we need, from the 
TVP, terms proportional to $\frac{\p \cdott \q}{\p^2 \q^2}$: they are obtained from the 
first term in (\ref{eq:mastac}): 
\be
\cVrrrr(\p,-\p-\r,\q,-\q+\r)=- \delta^{a_1',a_2'} \delta^{b_1,b_2} \delta^{b_3,b_4} 
\frac{\sqrt{2}\pi}{N_c^2-1} g^4 \w^4\frac{\p\cdott
\q}{\p^2 \q^2}.
\label{eq:lncrerpq}
\ee
Combining these expressions and performing the integrals we obtain:
\be
(K_1\,K_2)\otimes\cVrrrr=\delta^{a_1',a_2'} \delta^{b_1,b_2} \delta^{b_3,b_4} 
\frac{\sqrt{2}\pi}{N_c^2-1}
N_c^2\!\!\frac{g^8}{8(2\pi)^3}
\!\frac{\w^4}{3!}\!\!\!\left(\ln\frac{\w_0^2}{\w^2}\right)^3,
\label{eq:tot11}
\ee
where $|\!\w_0\!|$ is the momentum at the upper end of the BFKL kernels, 
specified by the condition
that it should be smaller than the momentum scales $|\l|$ and $|\m|$ 
which were considered in the collinear limit of the upper TPV. 
Convoluting this expression with the impact factor
$\imp_{\{b'\}}(\w)=\delta^{b_1'b_2'}\frac{2}{5}\B \frac{Q_0^2}{\w^2} 
\ln Q_0^2/\w^2$ below the vertex yields: 
\be
\begin{split}
(K_1\,K_2) \otimes\cVrrrr \otimes \imp_{\{a'\}} =\delta^{b_1,b_2}
 \delta^{b_3,b_4} 
\sqrt{2}\pi N_c^2
\frac{2}{5}\B \frac{\alpha_s^4}{2 \pi}
\frac{Q_0^2}{4!}\left(\ln\frac{\w_0^2}{Q_0^2}\right)^4.
\label{eq:kas1}
\end{split}
\ee
Here, in order to get the logarithmic contribution, we took, in (\ref{impactanticollinear}), 
the next-to-leading term in the 
anticollinear expansion of $\imp$ .
In the configuration $|\r| \gg|\q|\!\!\gg\!\!|\p|$ the same result is obtained.
\item  
Repeating a similar analysis in the case 
when $|\q|\!\!\gg\!|\p|\!\!\gg\!\!|\r|$, we find, from the last term on the 
rhs of (\ref{eq:mastac}):
\be
(K_1\,K_2) \otimes\cVrrrr \otimes \imp_{\{a'\}}
=\delta^{b_1,b_2} \delta^{b_3,b_4} 
\sqrt{2}\pi N_c^2
\frac{2}{5}\B \frac{2 \alpha_s^4}{\pi}
\frac{Q_0^2}{4!}\left(\ln\frac{\w_0^2}{Q_0^2}\right)^4.
\label{eq:cancel}
\ee
The same contribution is obtained from the region $|\p|\!\!\gg\!|\q|\!\!\gg\!\!|\r|$.
\item 
Finally, there are the regions  $|\q|\!\gg\!\!|\r|\gg\!\!|\p|$ and $|\p|\!\gg\!\!|\r|\!\gg\!|\q|$.
For the first case we use the first term in the second line of 
(\ref{eq:mastac}) and obtain:
\be
(K_1\,K_2) \otimes\cVrrrr \otimes \imp_{\{a'\}}
=\delta^{b_1,b_2} \delta^{b_3,b_4} 
\sqrt{2}\pi N_c^2
\frac{2}{5} \B \frac{\alpha_s^4}{\pi} 
\frac{Q_0^2}{4!}\left(\ln\frac{\w_0^2}{Q_0^2}\right)^4.
\label{eq:kas2}
\ee
The second region gives the same contribution.
\end{itemize}

\subsection{Virtual parts}
Let us now analyze the contribution coming from the virtual parts of the vertex
in the anticollinear limit. Again, we are looking for the maximal power 
of logarithms. We begin with the region  $|\r|\gg |\p|,|\q|$. Using (\ref{eq:vertex}),(\ref{eq:gvirt}) for the 
virtual parts of the TPV, (\ref{BFKLcoll}) for the BFKL kernel, (\ref{propagators})
for the propagators above the vertex, and (\ref{impactanticollinear}) for the lower impact factor 
we immediately see that none of the $G_2$ functions allows for three logarithmic integrals.
The same observation holds for the regions $|\q|\!\gg\!\!\!|\r|\gg\!\!|\p|$ and 
$|\p|\!\gg\!\!|\r|\gg\!|\q|$.
We are then left with the region $|\p|, |\q| \gg |\r|$. From the BFKL kernels and from the propagators 
we find the denominators $1/\p^2 \cdot 1/\q^2$, and we therefore need the factor $1/\r^2$ from the 
propagators below the TPV. They can come only from the first term in the function $G_2(1+2,3,4)$:
\be
g^2 G_2(1+2,3,4)= \alpha_s \frac{\w^4}{2\pi} \left(\ln \frac{\w_3^2}{(\w_3+\w_4)^2}\,\, \delta(\w+\w_3+\w_4)
            +\ln \frac{\w_3^2}{\w_4^2}\,\, \delta(\w+\w_4) \right),
\ee
and from analogous terms in $G_2(1+2,3,4)$, $G_2(1,2,3+4)$, $G_2(2,1,3+4)$. 
Convoluting these $G_2$ functions with the BFKL kernels and with the impact factor,
and setting the lowest momentum scale equal to $Q_0^2$, we find:
\be
(K_1\,K_2)\otimes \cVvv \otimes \imp_{\{a'\}}
=-\delta^{b_1,b_2} \delta^{b_3,b_4} 
\sqrt{2}\pi N_c^2\frac{2}{5} \B 
\frac{\alpha_s^4}{\pi}
\frac{Q_0^2}{2}\left(\ln\frac{\w_0^2}{Q_0^2}\right)^4.
\label{eq:cancel2}
\ee

Let us shortly summarize our results for the large-$N_c$ limit, before we continue the finite $N_c$ analysis.
Our goal was to find those terms of the twist expansion of the TPV which, after convolution with the BFKL kernel, would 
generate the maximal possible power of transverse momentum logarithms. In the
collinear case (upper TPV), we had to restrict the BFKL ladders below the 
TPV to the forward direction, and we therefore 
expected to find, from the $\m$ and $\l$ integrations, two logarithms. 
After the convolution with the upper impact factor, a third logarithm 
should appear. What we found is that the coefficient of this maximal 
number of logarithms vanishes, both 
for the connected and for the disconnected parts of the TPV.
In the anticollinear case we had to include the integral over the 
momentum transfer across the first BFKL kernel. After convoluting these 
integrals with the lower impact factor, we expect four logarithms. In fact, 
we found these logarithmic contributions, both in the real and in the 
virtual part of the TPV, and they came from different regions of ordered 
transverse momenta.   

This completes our twist-4 analysis of the one-loop Pomeron self-energy of the 
BFKL Pomeron (Fig.2b). We have found that the upper TPV vanishes at the twist-4 point,
whereas the lower one provides nonzero contributions.

\section{Finite $N_c$}
\subsection{The collinear limit}
In this section  we are going to investigate  contributions to the  vertex in (\ref{eq:kovl})
that are suppressed in the large $N_c$ limit.
Repeating our analysis of the previous sections we obtain for the first subleading piece:
\be
\cVrrr(1,3,2,4)^{\tau=4}=\nonumber
\ee
\be
= \delta^{a_1'a_2'}\delta^{b_1,b_2}\delta^{b_3,b_4}
\frac{\sqrt{2}\pi}{(N_c^2-1)^2}
g^4\; \k^2 \frac{4(- \k_1\cdott\k_3\k_2\cdott\k_4+ \k_1\cdott \k_2
\k_3\cdott\k_4+\k_1\cdott\k_4\k_2\cdott \k_3)}{\k^4}.
\ee
Substituting $\k_1=\l$, $\k_2=-\l$, $\k_3=\m$, $\k_4=-\m$ we obtain:
\be
\cVrrr(\l,\m,-\l,-\m)^{\tau=4}=\delta^{a_1'a_2'}\delta^{b_1,b_2}
\delta^{b_3,b_4}
\frac{\sqrt{2}\pi}{(N_c^2-1)^2}
g^4 \k^2
\frac{4\l^2\m^2}{\k^4}.
\ee
The convolution with the two BFKL kernels  gives:
\be
\csVrrr\otimes (K_1\,K_2) =
\delta^{a_1'a_2'} \delta^{b_1,b_2} \delta^{b_3,b_4}
\frac{\sqrt{2}\pi}{(N_c^2-1)^2} N_c^2 \frac{4 g^8}{(2\pi)^4}
\frac{\k_0^4}{\k^2} \left(\ln\frac{\k^2}{\k_0^2}
\right)^2.
\ee
Convolution with the impact factor gives:
\be
\impfa \csVrrr\otimes (K_1\,K_2)
=\frac{\sqrt{2}\pi}{N_c^2-1}
N_c^2\frac{2}{5}\B 
\frac{8 \alpha_s^4}{\pi^2}\frac{\k_0^4}{Q_1^4}
\frac{1}{3}\left(\ln\frac{Q_1^2}{k_0^2}\right)^3.
\label{eq:totsubreal}
\ee
The same result holds for the second subleading part.

For the virtual corrections to the TVP the situation is the same as 
for the leading-$N_c$ part: by simply inspecting the powers of 
transverse momenta, we find that the integrations over $\m$ and $\l$ are not 
logarithmic, i.e. they cannot generate the maximal power of logarithms.

\subsection{The anticollinear limit}
Here our starting expression for the real part of the TPV can be taken 
directly from the rhs of 
eq.(\ref{eq:mastac}), by interchanging $\w_2$ and $\w_3$. 
For the first nonleading piece we have:
\be
\cVrrr(\w_1,\w_3,\w_2,\w_4)^{\tau=-2}=\nonumber 
\ee 
\be
\begin{split}
\delta^{a_1',a_2'} \delta^{b_1,b_2} \delta^{b_3,b_4} 
\frac{\sqrt{2}\pi}{(N_c^2-1)^2} g^4 \w^4    
               \Bigg[-\frac{\w_1\cdott \w_2}{\w_1^2
\w_2^2}-\frac{\w_2\cdott\w_3}{\w_2^2 \w_3^2} -\frac{\w_1\cdott \w_4}{\w_1^2
\w_4^2}-\frac{\w_3\cdott \w_4}{\w_3^2 \w_4^2} \\
-\frac{\w_1\cdott (\w_1+\w_3)}{\w_1^2 (\w_1+\w_3)^2}-
                     \frac{\w_3\cdott (\w_1+\w_3)}{\w_3^2 (\w_1+\w_3)^2}-
                     \frac{\w_2\cdott (\w_2+\w_4)}{\w_2^2 (\w_2+\w_4)^2}-
                     \frac{\w_4\cdott (\w_2+\w_4)}{\w_4^2 (\w_2+\w_4)^2}\\
                   -\frac{(\w_1+\w_3)\cdott(\w_2+\w_4)}
                      {(\w_1+\w_3)^2(\w_2+\w_4)^2}\Bigg].
\end{split}
\label{eq:mastacsub}
\ee

The analysis is analogous to the leading-$N_c$ case. In detail we find:
\begin{itemize}
\item For $|\r|\!\!\gg\!\!|\p|\!\!\gg\!\!|\q|$ the logarithmic contribution 
comes, on the rhs of eq.(\ref{eq:mastacsub}), 
from the second term of the second line. We obtain:  
\be
(K_1\,K_2)\otimes\cVrrr \otimes \imp_{\{a'\}}=
\delta^{b_1,b_2}
 \delta^{b_3,b_4} 
\frac{\sqrt{2}\pi}{N_c^2-1} N_c^2
\frac{2}{5}\B \frac{\alpha_s^4}{2 \pi}
\frac{Q_0^2}{4!}\left(\ln\frac{\w_0^2}{Q_0^2}\right)^4.
\label{eq:anticolfiniNc1}
\ee
The same result holds for the region $|\r|\!\gg\!|\q|\!\gg\!|\p|$, taking
in eq.(\ref{eq:mastacsub}) the first term of the second line.
This result is same as in (\ref{eq:kas1}), except for the suppression 
by $N_c^2-1$.
\item $|\q|\!\!\gg\!\!|\r|\!\gg\!|\p|$: 
here we use the first term on the rhs of eq.(\ref{eq:mastacsub}) and obtain 
\be
(K_1\,K_2)\otimes \cVrrr \otimes \imp_{\{a'\}}
=\delta^{b_1,b_2} \delta^{b_3,b_4} 
\frac{\sqrt{2}\pi}{N_c^2-1} N_c^2
\frac{2}{5} \B \frac{\alpha_s^4}{\pi} 
\frac{Q_0^2}{4!}\left(\ln\frac{\w_0^2}{Q_0^2}\right)^4
\label{eq:anticolfiniNc2}.
\ee
The region $|\p|\!\!¸\gg\!\!|\r|\!\gg\!|\q|$ gives the same result.
It coincides with (\ref{eq:kas2}), but is suppressed by $N_c^2-1$.
\end{itemize}
The regions $|\q|\!\!\!\gg\!\!\!|\p|\!\!\!\gg\!\!\!|\r|$ and $|\p|\!\!\!\gg\!\!\!|\q|\!\!\!\gg\!\!\!|\r|$ 
do not contribute to the maximal number of logarithms.

Finally we come to the virtual parts of the $N_c$-suppressed parts of the TPV. 
Repeating the analysis, carried out for the virtual part of the leading-$N_c$ piece, we 
find no contribution to the maximal number of logarithms. The final result for the 
anticollinaer limit of the $N_c$-suppressed part of the TPV, therefore, is given 
by the real piece alone.   
\section{Nonlinear evolution equations}
\subsection{General evolution equations}
Let us now make some use of the TPV in QCD reggeon field theory.
To be definite let us consider deep inelastic scattering on a hadronic target 
(a single proton or a nucleus). We define color singlet $t$-channel states of 
$n$ reggeized gluons ($n$ even) in the Heisenberg picture which are labeled by color and 
momentum degrees of freedom:
\begin{eqnarray}
|n\rangle & =& \frac{1}{\sqrt{n!}} a^{\dagger}_{a_1}(\k_1)... a^{\dagger}_{a_n}(\k_n)|0 
\rangle \nonumber \\ 
& = & |\k_1,..\k_n;a_1,..a_n\rangle.
\label{reggeonstates}
\end{eqnarray}
The normalization is:
\be
[a_{a}(\k),a^{\dagger}_{a'}(\k')]=(2\pi)^3 \k^2 \delta(\k -\k') \delta_{aa'}
\ee
and
\be
\langle n|n'\rangle=\delta_{n'n}\frac{1}{n'!}
\sum_{\sigma(n)}\prod_{i=1}^{n'}\left((2\pi)^{3}\delta(\k_i-\k_i')\k_i^{2}
\delta^{a_ia_i^{'}}\right),
\ee
where the sum extends over the permutations of outgoing gluons.
The unity operator is given by:
\be
\sum_{n}|\!n\!\rangle\langle\!n\!|=\sum_{n}^{\infty}\prod_{i=1}^{n}\int
\frac{d^2\k_i}{(2\pi^3)}
\frac{1}{\k_i^{2}}|\!\k_1,..\k_n;a_1,..a_n\rangle
\langle\!\k_1,..\k_n;a_1,..a_n\!|,
\ee 
where the summation on the left hand-side includes also the integration 
over the continuous degrees of freedom.

We assume that the target state, at some initial rapidity, can be written 
as a superposition:
\be
|p\rangle=\sum_{n=1}^\infty\!c_{n}|\!n\rangle .
\ee
The rapidity evolution of this (color singlet) state is given by:
\be
e^{yH}|p\rangle=|p(y)\rangle,
\ee
The Hamiltonian consists of several pieces(Figs. 8-10):
\be
H=H_{2\rightarrow 2}+H_{2\rightarrow 4}+H_{4\rightarrow 2}+ 
H_{2\rightarrow 6}+H_{6\rightarrow 2}+ ...\,\,.
\label{eq:ham0}
\ee
The first term denotes the case where, inside the $n$ gluon state,
only one pair of gluons interacts, in the second term one pair splits into 
four gluons etc. The matrix elements of $H_{2\rightarrow 2}$ are expressed 
in terms of the BFKL Hamiltonian:
\begin{align}
\langle n'|H_{2\rightarrow
2}|n \rangle=\delta_{nn'} 
\sum_{i>j=1}^{n'} \bigg[ f_{a_i a_i'c}f_{c a_j'a_j}  \bigg\{
 K_{2 \to 2}(\k_i,\k_j;\k_i',\k_j') (2 \pi)^3 \delta(\k_i+\k_j - \k_i' -\k_j') 
 \nonumber \\ 
+ \left( \omega(\k_i) 
+ \omega(\k_j)\right) \k_i^2 \k_j^2 \delta(\k_i-\k_i') \delta(\k_j-\k_j') 
\bigg\} + (\k_i' \leftrightarrow  \k_j', a_i' \leftrightarrow  a_j') 
\bigg] \nonumber \\
\frac{1}{(n'-2)!} 
\sum_{\sigma(n'-2)}
\prod_{l\neq i,j}^{n'}(2\pi)^3 \delta(\k_l-\k_l')\delta_{a_l a_l'} 
\label{eq:ham1}
\end{align}
where the second term in the square bracket stands for the symmetrization of the outgoing 
two-gluon state. This kernel corresponds to the BKP interaction in the color singlet state.
All the other terms are presently known only for the special case where not only the total $n$ gluon system but also the interacting 
subsystem belongs to the color singlet representation.
\newpage
\begin{figure}[t!] \centerline{\epsfig{file=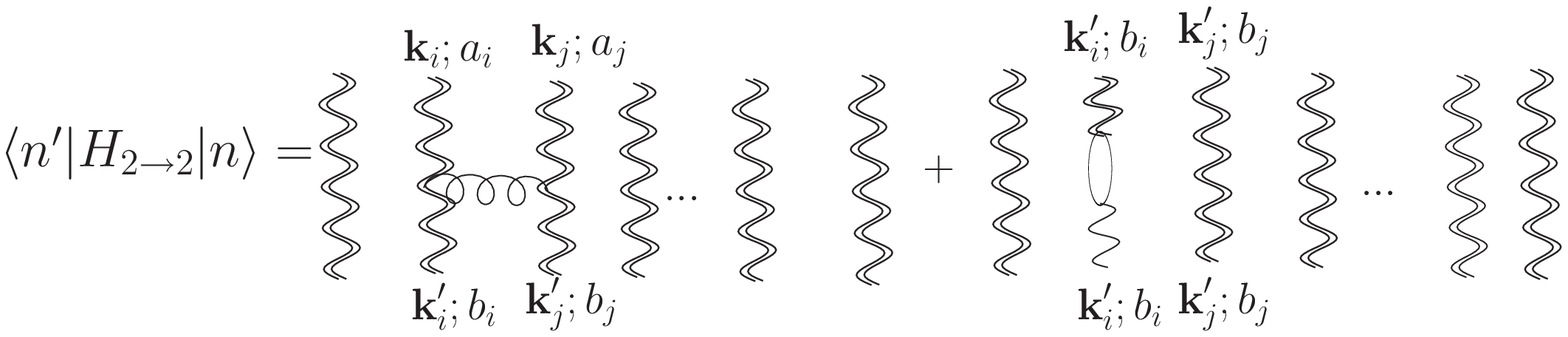,height=3.8cm}}
\caption{\em Matrix element given by eq.(\ref{eq:ham1}).}
\label{fig:hamm1}
\end{figure}
\vspace{1cm}
\begin{figure}[t!] \centerline{\epsfig{file=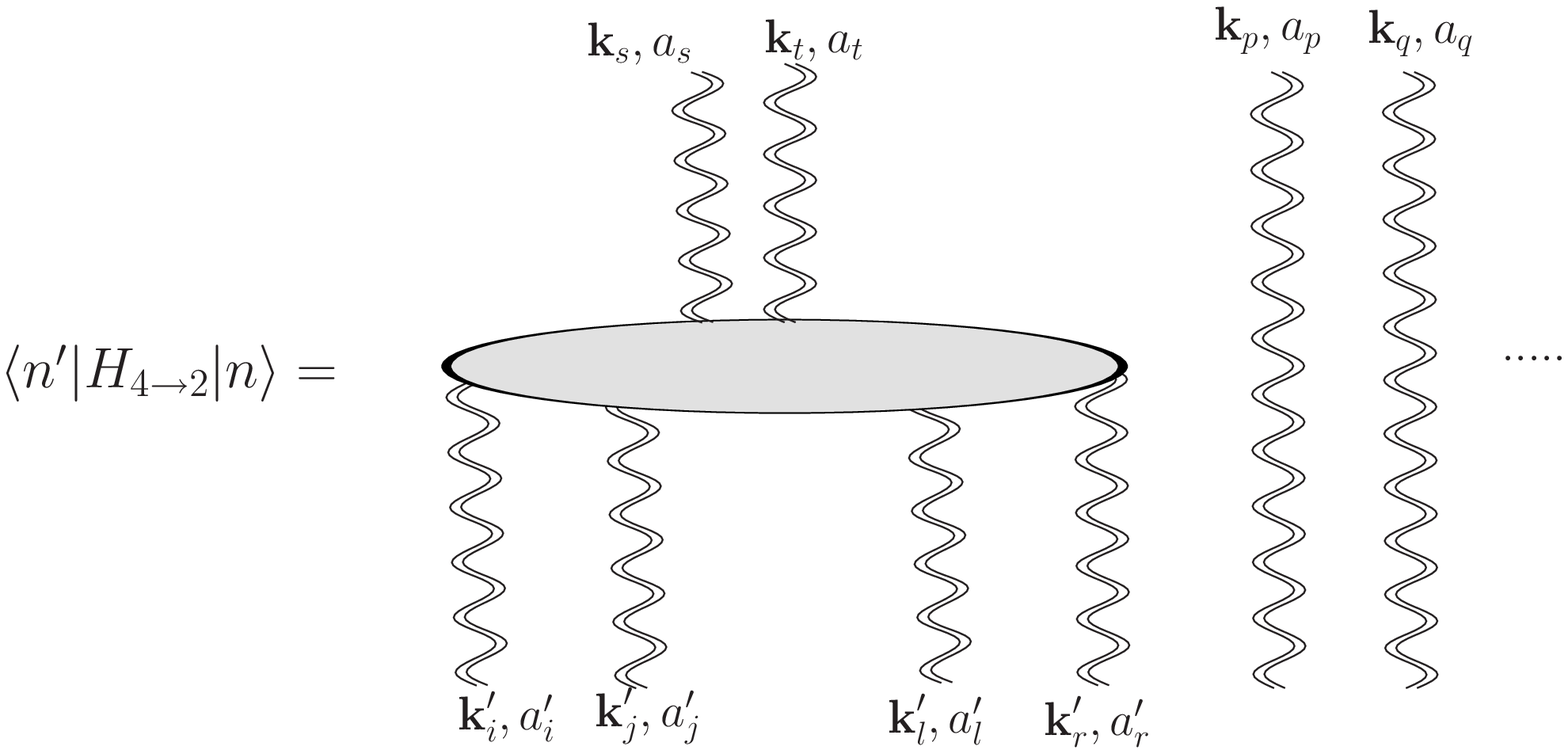,height=7cm}}
\caption{\em Matrix element given by eq.(\ref{eq:ham3}).}
\label{fig:}
\end{figure}
\begin{figure}[t!] \centerline{\epsfig{file=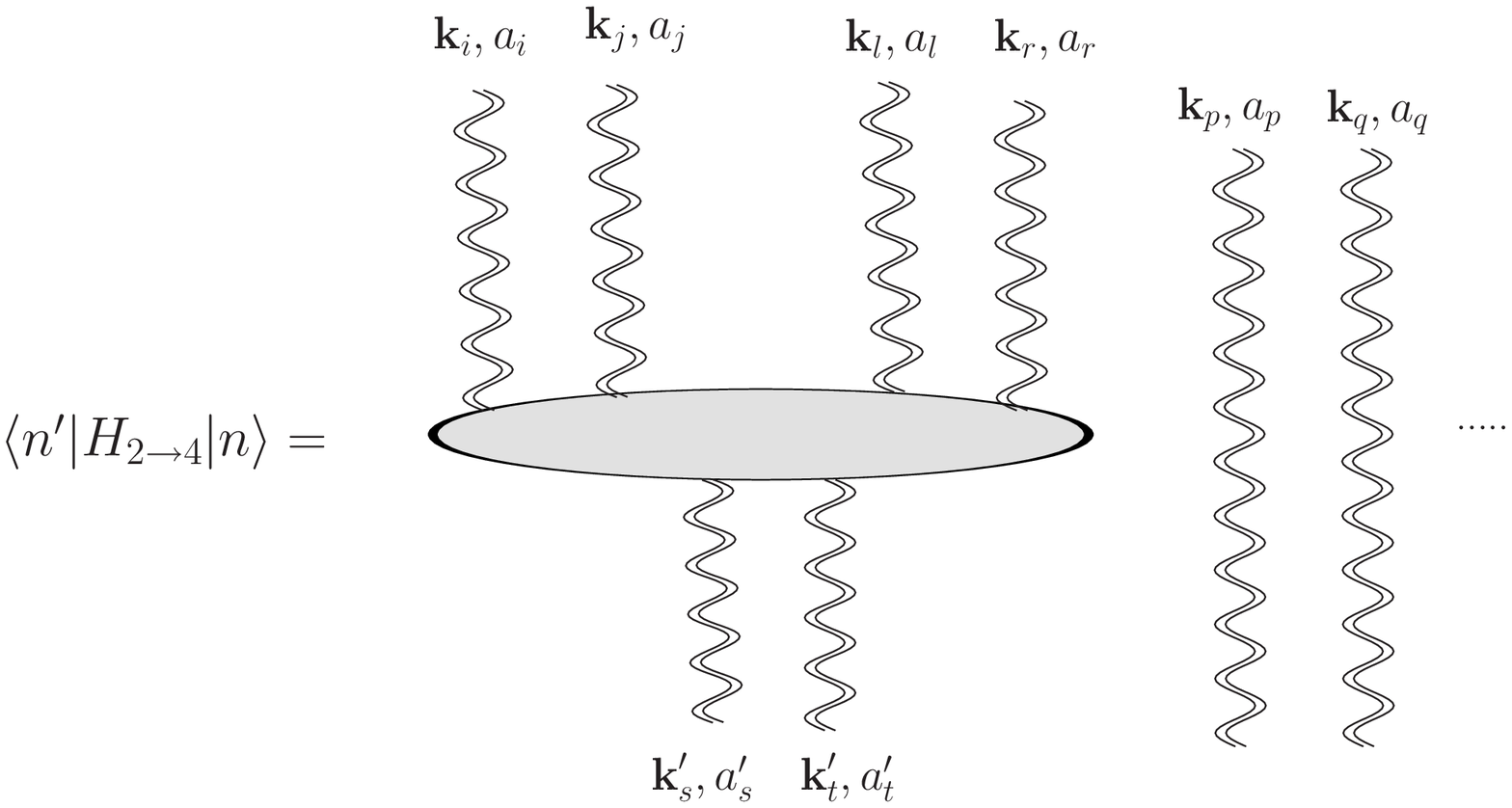,height=7cm}}
\caption{\em Matrix element given by eq.(\ref{eq:hamm2}).}
\label{fig:hamm2}
\end{figure}
\newpage
In particular,
the second term contains the $2 \to 4$ transition vertex:
\be
\begin{split}
\langle n'|H_{2\rightarrow 4}|n \rangle=\delta_{n\,n'+2}\sum_{s\!>\!t=1}^{n'}
\sum_{i\!>\!j\!>\!l\!>\!r\!=1}^{n}
\bigg[{\cal V}^{a_i' a_j' a_l' a_r';a_s,a_t}(\k_i',\k_j',\k_l',\k_r';\k_s,\k_t) +  
(\k_s \leftrightarrow \k_t, a_s \leftrightarrow a_t) \bigg]\\
(2 \pi)^3 \delta(\k_i+\k_j+\k_l+\k_r-\k_i-\k_j)
\frac{1}{(n-4)!} \sum_{\sigma(n-4)} \prod_{p\neq i,j,l,r}^{n} (2\pi)^3 \delta(\k_p-\k'_p)\delta^{a_pa'_p},
\end{split}
\label{eq:ham3}
\ee
whereas the third term allows for four gluon to fuse into two gluons:
\be
\begin{split}
\langle n'|H_{4\rightarrow 2}|n\rangle=
\delta_{n\,n'-2}\sum_{s\!>\!t}^{n}\!
\sum_{i\!>\!j\!>\!l\!>\!r\!=1}^{n'}\!\!\!\!\!\bigg[ {\cal V}^{a_sa_t;a_i'a_j'a_l'a_r'}
(\k_s',\k_t';\k_i,\k_j,\k_l,\k_r) + (\k_s' \leftrightarrow \k_t', a_s' \leftrightarrow a_t') \bigg]\\
(2 \pi )^3 \delta(\k_i+\k_j+\k_l+\k_r-\k_s'-\k_t')
\frac{1}{(n'-4)!} \sum_{\sigma(n'-4)}
\prod_{p\neq i,j,l,r}^{n'} (2\pi)^3 \delta(\k_p-\k'_p)\delta^{a_pa'_p}.
\label{eq:hamm2}
\end{split}
\ee
The next two terms on the rhs of (\ref{eq:ham0}) belong to the 
Pomeron $\to$ two 
Odderon vertex ~\cite{BaEwerz} (restricted to the color singlet channel) 
and to its inverse, resp. They will not be discussed further. 
Higher order kernels (indicated by the dots) 
have not been computed yet. 
Let us define the $n$ reggeon wave function component of the target at rapidity $y$ 
in the following way: 
\be
\psi_{n}^{\{a_i\}}(y,\k_1,\k_2...\k_n) =\langle n|e^{yH}|p\rangle .
\label{reggeoncomponent} 
\ee
Upon differentiation with respect to $y$ we obtain:
\bea
\frac{\partial \psi^{\{a_i\}}_{n}}{\partial y}=\langle
n|He^{yH}|p\rangle =\sum_{n'}\langle n|H|n'\rangle\langle
n'\!|e^{yH}|\!p\rangle\nonumber\\= \sum_{n'}\langle n| H|\!n'\rangle \psi^{\{a_i'\}}_{n'} .
\label{eq:coupledeq}
\eea
This defines an infinite set of coupled equations. 
It cannot be closed because, for instance, the equation for the two gluon 
wave function receives contributions coming from the four gluon wave function:
\be
\frac{\partial \psi^{a_1a_2}_2}{\partial y}=
\langle 2|H_{2\rightarrow 2}|2\rangle \psi_2^{a_1a_2}-
\langle2|H_{4\rightarrow 2}|4\rangle \psi_4^{a_1a_2}
\ee
(the term proportional to (\ref{eq:ham3}) vanishes since 
it requires zero gluons in the initial state). 
\subsection{The nonlinear equation for the unintegrated gluon density}
In order to reach further simplification, we take the large-$N_c$ limit. 
In practice this implies that 
we group the $n$ gluons into $n/2$ color singlet pairs (Pomerons) and 
associate with each pair a color singlet projector: this projector acts on 
the color tensors of the interaction Hamiltonians and leads to color weight 
factors of the interaction kernels. In particular, in the $2 \to 2$ 
Hamiltonian the color tensor $f_{a_ia_i'c} f_{ca_j'a_j}$ is replaced by the 
color factor $-N_c$, and in the $2 \to 4$ Hamiltonian,
$H_{2 \to 4}$, the $2 \to 4$ vertex ${\cal V}^{a_sa_t;a_i'a_j'a_m'a_n'}$ reduces to 
the function $V(\k_i',\k_j',\k_m',\k_n')$ (cf. eq.(\ref{eq:fvertex})). 
The evolution equations have to be 
reformulated in terms of $N$ states of gluon pairs: each pair carries 
two momentum variables, $\q$ and $\k$: $\q$ denotes the total tranverse 
momentum of the two gluon state, and $\k$, $\q-\k$ are the momenta of the two 
constituent gluons. The state consisting of $N=n/2$ such pairs is defined as:
\bea
|N\rangle & = &  \frac{1}{\sqrt{N!}} A^{\dagger}(\q_1,\k_1)...A^{\dagger}(\q_N,\k_N) |0\rangle
\nonumber \\
&=& |(\q_1,\k_1),..(\q_N,\k_N)\rangle     
\eea
(we use capital letters to distinguish the pair-basis from 
the reggeon basis). The normalization follows from 
\be
[A(\q,\k), A^{\dagger}(\q',\k')] =  (2\pi)^6 \k^2 (\q-\k)^2 \delta(\q - \q') \delta(\k - \k'),   
\ee
in analogy with the reggeon states. We write the Hamiltonian as
\be 
H = H_{1 \to 1} + H_{1 \to 2} +  H_{2 \to 1} , 
\ee  
where 
\bea
\langle1|H_{1 \to 1}|1\rangle = N_c \bigg\{
 K_{2 \to 2}(\k_i,\k_j;\k_i',\k_j') (2 \pi)^3 \delta(\k_i+\k_j - \k_i' -\k_j') 
 \nonumber \\ 
+ \left( \omega(\k_i) 
+ \omega(\k_j)\right) \k_i^2 \k_j^2 \delta(\k_i-\k_i') \delta(\k_j-\k_j') 
\bigg\} + (\k_i' \leftrightarrow  \k_j') 
\label{eq:ham2}
\eea
and
\bea
\langle1| H_{1 \to 2}|2\rangle=
\bigg[ V(\k_i',\k_j',\k_l',\k_r';\k_s,\k_t) +  
(\k_s \leftrightarrow \k_t, a_s \leftrightarrow a_t) \bigg]\\
(2 \pi)^3 \delta(\k_i+\k_j+\k_l+\k_r-\k_i-\k_j).
\eea
The amplitudes $\Psi_N$ in this basis of gluon pairs are defined in analogy 
with (\ref{reggeoncomponent}). 

Next we invoke the mean field approximation and make the following factorizing ansatz: 
\be
\Psi_2(y,\k_1,\q_1 -\k_1, \k_2,\q_2 - \k_2) =\Psi_1(y,\k_1,\q_1 -\k_1)
\Psi_1(y,\k_2,\q_2 - \k_2). 
\label{eq:factorisation}
\ee
This ansatz can be justified for a large nuclear target. It allows to obtain a closed 
equation for $\Psi_1$: 
\be
\frac{\partial \Psi_1}{\partial y}=
\langle 1|H_{1\rightarrow 1}|1\rangle \Psi_1- \frac{1}{\sqrt{2}} \langle 1|H_{2\rightarrow 1}|2\rangle
\Psi_1 \Psi_1.
\label{eq:balbk}
\ee 
To obtain the BK equation for the unintegrated gluon density let us define the off-diagonal unintegrated gluon 
density via:
\be
{\cal F}(y,\k_1,\k_2) = \Psi_1 (y,\k_1,\k_2) =\langle 1|e^{-yH}|p\rangle.
\label{eq:defgluonu}
\ee 
Using (\ref{eq:ham1}), (\ref{eq:ham3}),
(\ref{eq:ham2}), (\ref{eq:balbk}), (\ref{eq:defgluonu}) we obtain the nonlinear 
evolution equation:  
\be 
\begin{split}
\frac{\partial{\cal F}(x,\q,\k)}{\partial \ln1/x}=
\!\!\int\!\!\frac{d^2\l}{(2\pi)^3}K(\l,\q-\l;\k,\q-\k)\frac{{\cal
F}(x,\q,\l)}{\l^2(\q-\l)^2}\nonumber\\
\end{split}
\ee
\be
\begin{split}
-\pi\int d^2\r\frac{d^2\l}{(2\pi)^3}\frac{d^2\m}{(2\pi)^3}
V(\k,-\k+\q;\l,-\l-\frac{\q}{2}+\r,\m,-\m-\frac{\q}{2}-\r)\\
\times
\frac{{\cal F}(x,\frac{\q}{2}+\r,\l)}{\l^2(-\l+\frac{\q}{2}+\r)^2}
\frac{{\cal F}(x,\frac{\q}{2}-\r,\m)}{\m^2(-\m+\frac{\q}{2}-\r)^2}.
\label{eq:BKK}
\end{split}
\ee
The momenta entering the vertex from below are labeled by $\k_1'\!\!=\!\!\l$, 
$\k_2'\!\!=\!\!-\l\!-\!\q/2+\r$,
$\k_3'\!\!=\!\!\m$, $\k_4'\!\!=\!\!-\m-\q/2-\r$. The variable $\r$ 
stands for the loop momentum. In ~\cite{BLV} it has been shown that this 
equation coincides with the Balitsky-Kovchegov equation, provided 
the solutions ${\cal F}$ belong to the M\"obius class of functions 
(i.e. the Fourier transform vanishes when the two coordinates become 
identical). 
We make the assumption that the coupling to the proton goes via the form 
factor (with momentum transfer $\r$) 
\be
F(\r,R)= \frac{e^{\frac{-\r^2 R^2}{4}}}{2\pi},
\label{eq:formfactor}
\ee
(where $R$ has the meaning of the proton radius), and for  
${\cal F}(x,\r,\k)$ we make the ansatz:
\be
{\cal F}(x,\r,\k) =  {\cal F}(x,\k) F(\r,R).
\ee
Then the integration over $\r$ on the rhs of (\ref{eq:BKK}) will be restricted 
to small values $\r^2 \le 1/R^2$. Now we restrict ourselves to zero 
momentum transfer, $\q=0$, which corresponds to the integration over the 
impact parameter, and, 
as further approximation, we put $\r=0$ at the TPV:
in the dipole language, this means that the typical dipole size is assumed to be 
much smaller than the impact parameter $b$. This allows to carry out 
the $\r$ integral, and one easily sees that  
the function ${\cal F}(x,\k)$ satisfies the somewhat simpler equation:
\be
\begin{split}
\frac{\partial {\cal F}(x,\k)}{\partial \ln1/x}=
\int\frac{d^2\l}{(2\pi)^3}K(\l,-\l;\k,-\k)\frac{{\cal F}(x,\l)}{\l^4}\\
-\pi\frac{1}{2\pi R^2}
\int\frac{d^2\l}{(2\pi)^3}\frac{d^2\m}{(2\pi)^3} V(\k,-\k;\l,-\l,\m,-\m)
\frac{{\cal F}(x,\l)}{\l^4}\frac{{\cal F}(x,\m)}{\m^4}.
\label{eq:ffaneq}
\end{split}
\ee
In the next step we perform the integrations over the azimuthal angles of $\l$, $\m$, and $\k$. 
Denoting the integrated function 
${\cal F}(x,\k)$ by $f(x,\k^2)$:
\be
f(x,\k^2) = \frac{1}{2\pi}\int\!d\phi {\cal F}(x,\k)
\ee
with
\be
xg(x,\k^2) =  \int^{\k^2} \frac{d\k'^2}{\k'^2} f(x,\k'^2),
\label{ugluondensity}
\ee
and using our result (\ref{eq:melG2}) for the angular averaged TPV,
the nonlinear equation reads\cite{KKM01,KK03}:
\be
\begin{split}
\frac{\partial f(x,\k^2)}{\partial \ln 1/x}= \frac{N_c\alpha_s}
{\pi}\k^2\int_0^{\infty}\frac{d\l^2}{\l^2}
\bigg[\frac{f(x,\l^2)- f(x,\k^2)}{|\k^2-\l^2|}+ \frac{
f(x,\k^2)}{\sqrt{(4\l^4+\k^4)}}\bigg]\\
-\frac{\alpha_s^2}{2R^2} \Bigg\{2\k^2
\bigg[\int_{\k^2}^{\infty}\frac{d\l^2}{\l^4}f(x,\l^2)\bigg]^2 
+2\;f(x,\k^2)\int_{\k^2}^{\infty}\frac{d\l^2}{\l^4}\ln\left(\frac{\l^2}{\k^2}\right)f(x,\l^2)
\Bigg\}.
\label{eq:faneq}
\end{split}
\ee

When applying this equation to the scattering of a virtual photon on a
nucleus we return to the question raised at the end of the introduction, 
the question of the most dominant gluon configurations. In the DGLAP approach 
one has strong ordering in momentum, i.e  virtualities of gluons closer to
photon are larger than those closer to the target. 
In the nonlinear evolution equation one then would expect that, at the 
kernel of the nonlinear term, the upper momenta, $k$, should be larger than the lower ones,
$\l^2$ and $\m^2$.
However, making use of our results for the collinear limit of the TVP and of
the structure of the angular averaged vertex, we arrive at the somewhat surprising conclusion 
that the momenta are ordered in the opposite direction. In more physical terms, the 
recombination of two smaller gluons ends up in a larger gluon. This suppression of softer 
gluons below the nonlinear term may explain why, in numerical solutions of the angular 
averaged BK equation for the unintegrated gluon ~\cite{KK03}, the BFKL diffusion into the infrared region 
is absent.  
\section{Comparison with other equations} 
As we have mentioned before, the nonlinear equation (\ref{eq:BKK}) coincides with 
the Balitsky-Kovchegov equation. In ~\cite{BLV}, the Fourier transform of (\ref{eq:BKK}) has 
been computed, and it has been shown that, in the class of M\"obius functions, 
it agrees with the BK equation. 

Alternatively, one can start ~\cite{KKM01,KK03,KS} from the Balitsky-Kovchegov equation for the dipole 
scattering amplitude in coordinate space, 
and compute the Fourier transform to momentum space. 
The connection between the momentum space gluon distribution ${\cal F}(x,\q,\k)$ and the dipole scattering 
amplitude is:
\be
{\cal F}(x,\q,\k) = \frac{N_c}{4\alpha_s \pi^2}\k^2(\k-\q)^2\nabla^2_{k}
\int\frac{d^2\x_{0}}{2\pi}\int\frac{d^2\x_{1}}{2\pi}e^{i\k\cdot\x_{0}}e^{i(\q-\k)\cdot\x_{1}}
\frac{N(\x_{01},\b,x)}{\x_{01}^2},
\label{eq:fourphi}
\ee         
where $\x_{01}\!=\!\x_0-\x_1$, and $\b\!=\!(\x_1+\x_2)/2$ is the impact parameter. Our steps of approximation described after (\ref{eq:BKK}) are equivalent  
to the factorization ansatz 
\be
N({\x_{01}},{\bf b},x)=N(\x_{01},x) S(\b)
\ee
and to the assumption that, in the Balitsky-Kovchegov equation, all dipole sizes are much smaller 
than the impact parameter $b$. With these approximations one arrives, after angular 
averaging and integration over impact parameter $b$,  at the nonlinear equation (\ref{eq:faneq}). 
 
Returning, once more, to the issue of the twist expansion, we have 
to conclude that the nonlinear BK equation, when restricting to solutions 
with conformal spin $n=0$, receives all its contributions from 'anticollinear'
terms. This, in connection with corrections to 
the single-ladder approximation at small $x$, makes the usefulness of a twist 
expansion some what doubtful.

Let us finally comment on other versions of nonlinear evolution equations.
The first nonlinear evolution equation which was a milestone in physics of saturation 
is the Gribov-Levin-Ryskin, Mueller-Qiu (GLR-MQ) equation \cite{GLR}, \cite{MQ}
(eqn(2.41) in \cite{GLR}, and eqn.(30) in \cite{MQ}), obtained in the double-logarithmic 
approximation:   
\be
\frac{\partial^2xg(x,\k^2)}{\partial \ln(1/x)\partial \ln\k^2}=\frac{\alpha_s N_c}{\pi}xg(x,\k^2)-
C\frac{\alpha_s^2}{\k^2R^2}[xg(x,\k^2)]^2.
\ee
(the constant C is not the same in the two papers; 
however, for our discussion this is not
essential). This equation can be rewritten in terms of the unintegrated gluon density $f(x,\k^2)$:
\be
\frac{\partial f(x,\k^2)}{\partial \ln1/x}=\frac{N_c\alpha_s}{\pi}\int_{k_0^2}^{\k^2}\frac{d\l^2}{\l^2} 
f(x,\l^2)-
C\frac{\alpha_s^2}{\k^2R^2}\bigg[\int_{\k_0^2}^{\k^2}\frac{d\l^{2}}{\l^{2}}f(x,\l^{2})\bigg]^2
\label{eq:GLRMQ}
\ee
The linear term coincides with the BFKL kernel in the collinear approximation. The nonlinear term 
should be interpreted as the TPV at the collinear limit. Its physical interpretation would support 
the strong ordering (collinear) picture discussed at the end of the previous section: 
momenta above ($k^2$) are larger than below (${k'}^2$) the nonlinear interaction.    
Our analysis, however, does not agree with this form of the nonlinear term. 
The structure of integrals is totally different. In particular,
we have come to the conclusion that, after angular averaging, the TPV does 
not contribute to the collinear limit. 

The GLR paper \cite{GLR} also presents another nonlinear equation (eq.(2.108)),
derived from summing up, at small $x$, single logs of the fan diagrams.  
It is written directly for the unintegrated gluon density which, in the GLR notation, 
differs from our definition (eq.(\ref{ugluondensity}):
\be
xg(x,\k^2)=\int^{k^2}\!\!\!d\k'^2\Phi(x,\k'^2).
\ee
This equation is an attempt to generalize the BFKL equation 
to the physics of dense systems, and its form is quite close to our equation (\ref{eq:faneq}):
\be
\frac{\partial\Phi(x,\k^2)}{\partial\ln1/x}=\frac{N_c\alpha_s}{\pi}\int^{\infty}_0\frac{d\l^2}{\l^2}\left[\frac{\Phi(x,\l^2)-
\Phi(x,\k^2)}{|\l^2-\k^2|}-\frac{\Phi(x,\k^2)}{\sqrt{4\l^4+\k^4}}\right]
-g_{TPV}\frac{1}{4\pi R^2}\left(\frac{\alpha_s}{4\pi}\right)^2\Phi^{2}(x,\k^2),
\ee
where $g_{TPV}$ is the local approximation of the following TPV vertex: 
\be
V\otimes\left(\Phi(x,\l^2) \Phi(x,\m^2)\right)=\int\frac{d\m^2}{\l^2}\frac{d\l^2}{\l^2}
\alpha_s(\m^2)\alpha_s(\l^2)\Phi(x,\m^2)
\Phi(x,\l^2)\theta(\l^2-\k^2)
\theta(\m^2-\k^2).
\label{eq:vertglr}
\ee
This vertex contains the same $\theta$-functions as in (\ref{eq:melG2}), and it thus 
supports the physical picture describes at the end of the previous section.
On the other hand, the detailed analytic form of the vertex is different from (\ref{eq:faneq});
in particular, it does not contain 
the disconnected pieces which, in the original derivation of the $2 \to 4 $ vertex, 
can be traced back to the reggeization of the gluon (there are also differences in the prefactors).

Despite these differences in the detailed form of the nonlinear equations  
it may very well be that, as far as the gross features of saturation are concerned, 
the qualitative behavior of solutions will be similar. 
It would be interesting to study this in more detail.

\section{Conclusions} 
In this paper we have investigated the momentum space triple Pomeron vertex. In particular, we 
have studied its collinear and anticollinear limits. This question arises naturally if 
one studies nonlinear corrections to the linear BFKL evolution in deep inelastic scattering at small $x$: 
one expects that, at least on the average, transverse momenta decrease when moving from the 
photon to the proton. In a first step one is then led to consider the limit of strong ordering.
Restricting ouselves to solutions with conformal spin zero, we have shown, 
for the simplest example of a fan diagram with one triple Pomeron vertex in the 
large-$N_c$ limit, that there is no contribution from the configuration of 
strongly ordered gluons. Beyond the large-$N_c$ limit such contributions exist. 

We have also constructed a set of evolution equations for the interaction of 
a photon with a nuclear target,
which, in the mean field approximation, reduces to a nonlinear evolution 
equation for the skewed unintegrated gluon density which, in the forward 
region, agrees with equation obtained in \cite{KKM01,KK03}.
We have also compared our momentum 
space expression for the nonlinear evolution kernel with different other versions discussed 
in the literature. We agree with the BK equation, but we find 
disagreement with other earlier versions of nonlinear evolution equations. 

Interpretating our results in terms of twist, we have shown that the BK-equation, 
when restricted to solutions with conformal spin zero, receives all its 
contributions from 'anticollinear' configurations, quite in contrast to 
the expected ordering of transverse momenta.    

We also hope that our analysis will help to analyse further
the contributions of pomeron loops.
\newpage 
\section*{Acknowledgments}
Useful discussions with Krzysztof
Golec-Biernat, Yuri Kovchegov, Lev Lipatov,
Leszek Motyka, Al Mueller, Agustin Sabio-Vera, and 
Michele Salvadore are gratefully acknowledged.
During this research Krzysztof Kutak has been supported by the Graduiertenkolleg 
{\it ``Zuk\"unftige Entwicklungen in der Teilchenphysik''}.

\end{document}